\documentclass[iop,apj,tighten]{emulateapj}
\usepackage{hyperref}

\journalinfo{The Astrophysical Journal, 722:65-78, 2010 October 10} 
\slugcomment{Received 2010 March 17; accepted 2010 July 19; published 2010 September 16}

\shorttitle{THE ORIGIN OF TURBULENCE IN CORONAL LOOPS}
\shortauthors{Rappazzo, Velli, \& Einaudi}

\begin{document}

\title{Shear Photospheric Forcing and the Origin of Turbulence in Coronal Loops}

\author{A. F. Rappazzo$^{1,2}$, M. Velli$^{2,3}$, and G.~Einaudi$^4$}
\affil{\vspace{.2em}
$^1$Instituto de Astrof\'{\i}sica de Canarias, 38205 La Laguna, Tenerife, Spain;  
\href{mailto:rappazzo@iac.es}{rappazzo@iac.es}\\ 
\vspace{.2em}
$^2$Jet Propulsion Laboratory, California Institute of Technology, Pasadena, CA 91109, USA\\
\vspace{.2em}
$^3$Dipartimento di Fisica e Astronomia, Universit\`a di Firenze, 50125 Florence, Italy\\
\vspace{.2em}
$^4$Dipartimento di Fisica ``E.~Fermi'', Universit\`a di Pisa, 56127 Pisa, Italy}

\begin{abstract}
We present a series of numerical simulations aimed at understanding the 
nature and origin of turbulence in coronal loops in the framework of the
Parker model for coronal heating.
A coronal loop is studied via reduced magnetohydrodynamics simulations in 
Cartesian geometry. A uniform and strong magnetic field threads the volume
between the two photospheric planes, where a velocity field in the form 
of a 1D shear flow pattern is present.
Initially  the magnetic field that develops in the coronal loop is a simple map of the
photospheric velocity field. This initial configuration is unstable to a multiple tearing 
instability that develops islands with $X$ and $O$ points in the plane orthogonal 
to the axial field.
Once the nonlinear stage sets in the system evolution is characterized by a 
regime of MHD turbulence dominated by magnetic energy.  A well developed 
power law in energy spectra is observed  
and the magnetic field never returns to the simple initial 
state mapping the photospheric flow. The formation of $X$ and $O$ points in the planes orthogonal
to the axial field allows the continued and repeated formation and dissipation of small scale current sheets
where the plasma is heated. We conclude that the observed turbulent dynamics
are not induced by the complexity of the pattern that the magnetic field-lines footpoints follow 
but they rather stem from the inherent nonlinear nature of the system.\\[.5em]
\textit{Key words:} magnetohydrodynamics (MHD) --- Sun: corona --- Sun: magnetic topology --- turbulence\\[.5em]
\textit{Online-only material}: animations at 
\href{http://www.df.unipi.it/~rappazzo/shear/}{http://www.df.unipi.it/$\scriptstyle{\sim}$rappazzo/shear/} 
(\textit{non permanent link})\\
\end{abstract}

\setcounter{page}{65}

\section{Introduction}

In recent papers \citep{rved07,rved08} we have described reduced magnetohydrodynamics
(RMHD) simulations of the Parker problem \citep{park72,park88,park94} for coronal loops in 
Cartesian geometry. We have shown that the system develops small scales, organized in current sheets
elongated in the direction of the DC magnetic field, through an MHD turbulent
cascade, and that a well defined power law spectrum is developed for total energy. 
The energy spectra  develop  steep slopes [seen also in the similar simulations of a line-tied RMHD system
by  \cite{dgm03}] with spectral indices  going from the  classical $-5/3$ Kolmogorov spectrum up to almost~$-3$.
The energy spectral indices (slopes) have a bearing on the heating rate of the boundary-forced system. The heating rates 
are  significantly increased for realistic values of  the magnetic field intensity and loop length
compared to the scaling laws with fixed indices \citep{dg99}, the last  being recovered when a Kolmogorov-like spectrum 
develops (i.e.\ for weak fields or long loops). 

In the published simulations we have always used photospheric velocity patterns
made up of large spatial scale projected convection cells (\emph{large-scale eddies}), 
mimicking disordered photospheric motions. As a consequence the magnetic field developing in the corona
was not in equilibrium, but dynamical evolution occurred from the outset with a time-scale
set by the interplay of forcing and nonlinearity.
In this paper, in order to clarify the origin of the MHD turbulent
dynamics found in our previous work, we explore the dynamics of coronal loops
system when a shear velocity flow stirs the footpoints of the magnetic field-lines.
It is  commonly thought that the topology of the photospheric driver
should strongly influence the dynamics of a coronal loop,
and that the magnetic field-lines anchored to the photospheric planes
should \emph{passively} follow their footpoints motions.
In this picture the electric currents should develop along neighboring
field-lines whose footpoints have a relative shear motion.
In particular it might be argued that  in our previous simulations the turbulent dynamics 
of the coronal medium originated from the ``complexity'' of the photospheric forcing patterns, 
with their large-scale eddies and stagnation points.

The \emph{simple} unidimensional shear forcing velocity 
used here allows  a very \emph{clear-cut numerical experiment}: 
if the field-lines were to passively follow the imposed footpoint motion only a sheared
magnetic field would develop inside the volume. 
The dynamics of these current layers are subject to tearing instabilities \citep{fkr63}.
If this were the physical process at work the topology of the magnetic field
would remain a \emph{mapping} of the forcing velocity pattern,  periodically disrupted by 
tearing-like instabilities when the shear grew beyond a threshold amount.
Simulations show that initially the magnetic field is sheared and a tearing instability
develops, but afterwards turbulent dynamics similar to those with more \emph{complex} 
boundary patterns develop, clarifying that turbulence does not stem from a direct
influence of the photospheric velocity pattern but it is due to the inherent nonlinear 
properties of the system.

The Parker Scenario for coronal heating has been the subject of
intense research. Both analytical \citep{su81,vb86,ant87,ber91,gff92,cls97,nb98,uz07,low09, 
lj09,bat09,aa10} 
and numerical \citep{msvh89,ls94,hvh96,gn96,hz09, hbz10} investigations have been carried out.
Given the complexity of  such boundary forced field-line tangling systems
simplified models have been developed, including 2D incompressible MHD models with magnetic forcing 
\citep{evpp96,dg97,gve98,dgd98,ev99} and shell models \citep{nmcv04,bv07}. 
Insights gleaned from these models include the important result that dissipation in forced MHD turbulence
occurs in the form of ``bursty'' events with well defined power law distributions in energy release, peak dissipation, and 
duration, in a way reminiscent of, and consistent with, the distribution of flares in the solar corona.

An analytical model of a forced system very similar to the simulation presented here
was proposed by \cite{hp92}, and recently extended to the anisotropic turbulence regime
by \cite{bgp08}. They started with
same MHD system threaded by a strong axial magnetic field in 
cartesian geometry and apply at the top and bottom boundaries 
two 1D velocity fields of opposite direction and assumed that the sheared structure that develops
in the corona then dissipates via an effective ``turbulent resistivity'' provided by a cascade,
so that a dissipative equilibrium is set up in which shearing is balanced by slippage provided
by the turbulence. This amounts essentially to a \emph{one-point closure} model of MHD turbulence
\citep{bis03}, where turbulence acts only on very small-scales while
the large-scales remain laminar and indeed with the same large-scale magnetic structure.
They  then use the eddy-damped quasi-normal Markovian 
approximation (EDQNM) \citep{pfl76} to estimate the effective cascade and dissipation for the given
driving shear, and this allow them to develop a heating theory in which the only free parameter is
the equivalent Kolmogorov constant. In contrast, our simulations will show that nonlinearity cannot be neglected 
even at the large-scales, so that the turbulent self-consistent state has little resemblance to the imposed shear flow.
As a result \cite{hp92} overestimate the heating rate as laminar dynamics
would lead to a higher energy injection (Poynting flux).

More recently \cite{dka05,dlkn09} have proposed the so-called ``secondary
instability'' as a leading mechanism operating in the Parker Scenario,
responsible for the rapid release of energy. In their view, disruption on ideal time-scale
must arise after some time while slow quasi-steady reconnection allows magnetic energy to continue to accumulate in the 
system.  In their view the system evolution may be
described by a sequence of equilibria destabilized by magnetic reconnection.
The bulk of numerical simulations performed by 
\cite{evpp96,dg97,gve98,dg99,ev99,dgm03,rved07,rved08}
has proven that the system does not evolve through a sequence of equilibria,
rather more complex dynamics develop. 

The initial setup of the simulation presented in \citep{dlkn09} is also very similar to the one
implemented in this paper but, besides the lower resolution and therefore the higher
influence of numerical diffusion, 
the time interval for which they advance the equations 
is too short compared with coronal loops and active region time-scales. 
These leads them 
to claim as representative of the dynamics what is  actually a \emph{transient} 
event taking place only during the early stage of the dynamics, in our case the multiple 
tearing mode current sheet 
collapse. This evolution is not generic, but is only representative of a small class of very symmetric 
boundary velocity patterns, those which admit coronal equilibria at all times. We will return to this question
and a more detailed discussion in the conclusion.

The paper is organized as follows. In \S~\ref{sec:gebc} we describe the basic governing equations 
and boundary conditions, as well as the numerical code used to integrate them.
In \S~\ref{par3} we discuss the initial conditions for our simulations and briefly summarize
the linear stage dynamics more extensively detailed in \cite{rved08}, while in
\S~\ref{sec:ed} we outline the main points of \cite{hp92} relevant to this work.
The results of our numerical simulations are presented in \S~\ref{sec:ns}, while
the final section is devoted to our conclusions and discussion of the
impact of this work on coronal physics.

\section{Governing Equations and Boundary Conditions} \label{sec:gebc}

We model a coronal loop as an axially elongated Cartesian box with an orthogonal
cross section  of size $\ell$  and an axial length $L$  embedded in an homogeneous 
and uniform axial magnetic field $\mathbf{B_0} = B_0\, \mathbf{\hat{e}_z}$ aligned along
the $z$-direction. Any curvature effect is neglected.

The top and bottom plates ($z=0$ and $L$) represent the photospheric surfaces where
we impose,  as boundary conditions, velocity patterns mimicking photospheric motions.
Along the $x$ and $y$ directions periodic boundary conditions are implemented.

At the top plate $z=L$ we impose a sinusoidal shear flow with wavenumber~$4$
\begin{equation} \label{eq:f0}
\mathbf{u^L} \left( x, y \right)  = 
  \sin \left( 4\, \frac{2\pi}{\ell}\,  x +1 \right) \, \mathbf{\hat{e}_y}.
\end{equation}
At the bottom plate $z=0$ we generally impose a vanishing velocity
\begin{equation} \label{eq:f1}
\mathbf{u^0} \left( x, y \right)  =  0,
\end{equation}
except in one simulation [run~F (table~\ref{tbl})] where, in order to compare with 
previous simulations implementing a vortical velocity forcing applied at both 
plates \citep{rved07,rved08}, the reversed pattern of the top 
plate~(\ref{eq:f0}) is applied 
\begin{equation} \label{eq:f2}
\mathbf{u^0} \left( x, y \right)   =  
- \sin \left( 4\, \frac{2\pi}{\ell}\,  x +1 \right) \, \mathbf{\hat{e}_y}.
\end{equation}
Here $\mathbf{\hat{e}_y}$ is the unitary vector directed along the $y$ direction, while
the flow is sheared along $x$.

The dynamics are integrated, as in our previous works, using the equations
of RMHD \citep{kp74,str76,mon82}, which are well suited for a plasma embedded in a strong 
axial magnetic field. In dimensionless form they are given by:
{\setlength\arraycolsep{-10pt}
\begin{eqnarray}
&&\frac{\partial \mathbf{u_{_\perp}}}{\partial t}  + 
\left(  \mathbf{u_{_\perp}} \cdot \nabla_{_{\!\perp}} \right) \mathbf{u_{_\perp}} = 
- \nabla_{_{\!\perp}} \left( p + \frac{ \mathbf{b_{_\perp}}^2 }{2} \right)
\nonumber \\
&& \qquad + \left(  \mathbf{b_{_\perp}} \cdot \nabla_{_{\!\perp}} \right) \mathbf{b_{_\perp}}
+ c_{A}\, \frac{\partial \mathbf{b_{_\perp}}}{\partial z}
+ \frac{(-1)^{n+1}}{Re_n} \nabla_{_{\!\perp}}^{2n}\, \mathbf{u_{_\perp}}, 
\label{eq:eq1} \\
&&\frac{\partial \mathbf{b_{_\perp}}}{\partial t}  + 
\left(  \mathbf{u_{_\perp}} \cdot \nabla_{_{\!\perp}} \right) \mathbf{b_{_\perp}} = 
\left(  \mathbf{b_{_\perp}} \cdot \nabla_{_{\!\perp}} \right) \mathbf{u_{_\perp}} 
+ c_{A}\, \frac{\partial \mathbf{u_{_\perp}}}{\partial z}
\nonumber \\
&& \qquad \qquad \qquad \qquad \qquad \qquad \qquad 
+ \frac{(-1)^{n+1}}{Re_n} \nabla_{_{\!\perp}}^{2n}\, \mathbf{b_{_\perp}}, 
\label{eq:eq2} \\
&& \nabla_{_{\!\perp}} \cdot \mathbf{u_{_\perp}} = 0, \qquad 
 \nabla_{_{\!\perp}} \cdot \mathbf{b_{_\perp}} = 0,
\label{eq:eq3} 
\end{eqnarray}
}where $\mathbf{u_{_\perp}}$ and  $\mathbf{b_{_\perp}}$ are the velocity and magnetic
fields components orthogonal to the axial field, $p$ is the kinetic pressure.
The gradient operator has components only in the perpendicular $x$-$y$ planes
\begin{equation}
\nabla_{_{\!\perp}} = \mathbf{\hat{e}_x} \frac{\partial}{\partial x} + 
\mathbf{\hat{e}_y} \frac{\partial}{\partial y}
\end{equation}
while the linear term $\propto \partial_z$ couples the planes along
the axial direction through  a wave-like propagation at the Alfv\'en 
speed $c_A$.
Incompressibility in RMHD equations follows from the large value
of the axial magnetic fields  \citep{str76} and they remain valid
also for low $\beta$ systems \citep{zm92,bns98} such as the corona.

To render the equations nondimensional, we have first expressed
the magnetic field as an Alfv\'en velocity [$b \rightarrow b/\sqrt{4\pi \rho_0}$],
where $\rho_0$ is the density supposed homogeneous and constant,
and then all velocities have been normalized to the velocity
$u^{\ast} = 1\ km\, s^{-1}$,  the order of magnitude of photospheric 
convective motions.

Lengths and times are expressed in units of the perpendicular length 
of the computational box $\ell^{\ast} = \ell$ and its related crossing time
$t^{\ast} = \ell^{\ast}/u^{\ast}$.
As a result, the linear terms $\propto \partial_z$ are multiplied by the
dimensionless Alfv\'en velocity $c_A = v_A/u^{\ast}$, where 
$v_A = B_0/\sqrt{4\pi \rho_0}$ is the Alfv\'en velocity associated with the axial magnetic field.

The majority of the simulations performed [specifically runs~A--E 
(see table~\ref{tbl})]  use a standard simplified diffusion model, 
in which both the magnetic resistivity $\eta$  and viscosity $\nu$ are constant 
and uniform. The kinetic and magnetic Reynolds numbers are
then given by:
\begin{equation}
Re = \frac{\rho_0\, \ell^{\ast} u^{\ast}}{\nu}, \qquad 
Re_{_m} = \frac{4\pi\, \ell^{\ast} u^{\ast}}{\eta c^2},
\end{equation}
where $c$ is the speed of light, and numerically they are given
the same value $Re=Re_{_m}$.
In equations~(\ref{eq:eq1})-(\ref{eq:eq2}) this case is realized for
$n=1$ with $Re_{_1} = Re$.

The index $n$ is called \emph{dissipativity} and for $n > 1$ 
the dissipative terms in (\ref{eq:eq1})-(\ref{eq:eq2})
correspond to so-called hyperdiffusion \citep{bis03}.
We use hyperdiffusion, with $n=4$, only in runs~F and G (table~\ref{tbl})
dedicated to study the energy spectra.
Hyperdiffusion is used because, even with a grid of $512 \times 512$ points 
in the x-y plane (the highest resolution grid we used for the plane),
the timescales associated with ordinary diffusion are small enough to affect 
the large-scale dynamics and render difficult the resolution of an 
inertial range. The diffusive time $\tau_{_n}$ at the scale $\lambda$ 
associated with the dissipative terms used in 
equations~(\ref{eq:eq1})-(\ref{eq:eq2}) is given by
\begin{equation}
\tau_{_n} \sim Re_{_n}\, \lambda^{2n}.
\end{equation}
For $n=1$ the diffusive time decreases relatively slowly toward smaller scales, 
while for $n=4$ it decreases far more rapidly. As a result for $n=4$ we have 
longer diffusive timescales at large spatial scales and diffusive timescales 
similar to the case with $n=1$ at the resolution scale. 
Numerically we require the diffusion time at the 
resolution scale $\lambda_{min} = 1/N$, where N is the number of grid points, to be of 
the same order of magnitude for both normal and hyperdiffusion, i.e.,
\begin{equation}
\frac{Re_{_1}}{N^2} \sim \frac{Re_{_n}}{N^{2n}} \quad
\longrightarrow \quad
Re_{_n} \sim Re_{_1}\, N^{2(n-1)}.
\end{equation}
Then for a numerical grid with $N=512$ points that requires
a Reynolds number $Re_{_1} = 800$ with ordinary diffusion we can 
implement $Re_{_{4}} \sim 10^{19}$ (table~\ref{tbl}), removing diffusive effects at the 
large scales and allowing (if present) the resolution of an inertial range.

We solve numerically equations~(\ref{eq:eq1})-(\ref{eq:eq3}) 
written in terms of the potentials of the orthogonal
velocity and magnetic fields [see \cite{rved07,rved08} for a 
more detailed description of the numerical code]
in Fourier space, i.e.\ we advance the Fourier components 
in the $x$- and $y$-directions of the scalar potentials. Along 
the $z$-direction, no Fourier transform is performed 
so that we can impose non-periodic boundary conditions (\S~\ref{par3}), 
and a central second-order finite-difference
scheme is used. In the $x$-$y$ plane, a Fourier pseudospectral method 
is implemented. Time is discretized with a third-order Runge-Kutta method.

\section{Initial Conditions and Linear Stage}  \label{par3}

At time $t=0$ we start our simulations with a uniform and homogeneous magnetic field
along the axial direction $\mathbf{B} = B_0\, \mathbf{\hat{e}_z}$. 
The orthogonal component of the velocity and magnetic fields are zero
inside our computational box $\mathbf{u_{_\perp}}=\mathbf{b_{_\perp}}=0$,
while at the top and bottom planes a large-scale velocity pattern is imposed 
[(\ref{eq:f0})-(\ref{eq:f1}) or (\ref{eq:f0})-(\ref{eq:f2})]  and kept \emph{constant in time}. 

We briefly summarize and extend to the shear forcing considered in this paper
the linear stage analysis covered in more detail
in \cite{rved08}. In general for an initial interval of time smaller than the 
nonlinear timescale $t < \tau_{nl}$,  nonlinear terms in equations 
(\ref{eq:eq1})-(\ref{eq:eq3}) can be neglected and the equations linearized. 
For simplicity we will at first neglect also the diffusive terms and consider
their effect in the second part of this section. 
The solution during the linear stage for generic boundary velocity forcings,
$\mathbf{u^L}$ and $\mathbf{u^0}$ respectively at the top and bottom planes
$z=L$ and $0$, is given by:
{\setlength\arraycolsep{-10pt}
\begin{eqnarray}
&&\mathbf{b_{_\perp}} (x,y,z,t) = 
\left[ \mathbf{u^L} (x,y) - \mathbf{u^0} (x,y) \right]
\frac{t}{\tau_A}, 
\label{eq:lin1} \\
&&\mathbf{u_{_\perp}} (x,y,z,t) = 
\mathbf{u^L} (x,y)\, \frac{z}{L} 
+ \mathbf{u^0} (x,y) \left( 1 - \frac{z}{L} \right),
\label{eq:lin2}
\end{eqnarray}
}where $\tau_A = L/v_A$ is the Alfv\'en crossing time along the axial direction $z$.
The magnetic field grows linearly in time, while the velocity field
is stationary and the order of magnitude of its rms is determined
by the boundary velocity profiles.
Both are linear combinations
(\emph{mapping}) of the boundary velocity fields.

Considering the boundary conditions (\ref{eq:f0})-(\ref{eq:f1}) imposed 
in most of the following simulations we obtain
{\setlength\arraycolsep{-10pt}
\begin{eqnarray}
&&\mathbf{b_{_\perp}} (x,y,z,t) = 
\frac{t}{\tau_A}\, \sin \left( 4\, \frac{2\pi}{\ell}\,  x +1 \right) \, \mathbf{\hat{e}_y}, 
\label{eq:lin1s} \\
&&\mathbf{u_{_\perp}} (x,y,z,t) = 
\frac{z}{L}\, \sin \left( 4\, \frac{2\pi}{\ell}\,  x +1 \right) \, \mathbf{\hat{e}_y}.
\label{eq:lin2s}
\end{eqnarray}
}Both fields are a clear \emph{mapping} of the shear velocity at the boundary,
with the magnetic field increasing its magnitude linearly in time.

When we shear the field-lines from both photospheric plates ($z=0$ and $L$)  using
the forcing (\ref{eq:f0})-(\ref{eq:f2}) we obtain a very similar result, still a \emph{mapping}
but with different amplitudes:
{\setlength\arraycolsep{-10pt}
\begin{eqnarray}
&&\mathbf{b_{_\perp}} (x,y,z,t) = 
2\, \frac{t}{\tau_A}\, \sin \left( 4\, \frac{2\pi}{\ell}\,  x +1 \right) \, \mathbf{\hat{e}_y}, 
\label{eq:lin1b} \\
&&\mathbf{u_{_\perp}} (x,y,z,t) = 
\left( 2\frac{z}{L} -1 \right) \, \sin \left( 4\, \frac{2\pi}{\ell}\,  x +1 \right) \, \mathbf{\hat{e}_y}.
\label{eq:lin2b}
\end{eqnarray}
}

For a generic forcing the solution  (\ref{eq:lin1})-(\ref{eq:lin2}) 
is valid only during the linear stage, while for $t > \tau_{nl}$ when the fields 
are big enough the nonlinear terms cannot be neglected.

Nevertheless there is a singular subset of velocity forcing
patterns for which the generated coronal fields (\ref{eq:lin1})-(\ref{eq:lin2})
have a vanishing Lorentz force and the nonlinear terms vanish exactly.
This subset  of patterns is characterize by having the \emph{vorticity constant
along the streamlines} \citep{rved08}.
In this case the solutions (\ref{eq:lin1})-(\ref{eq:lin2}) are an \emph{exact solution
at all times}, not an approximation valid only for $t < \tau_{nl}$.

As can be proved by direct substitution the sheared forcing (\ref{eq:f0})-(\ref{eq:f1})
[and also (\ref{eq:f0})-(\ref{eq:f2})] is one of these degenerate patterns, and the 
solution (\ref{eq:lin1s})-(\ref{eq:lin2s}) [or (\ref{eq:lin1b})-(\ref{eq:lin2b})] is exact at 
all times. 

So far we have neglected the diffusive terms in the RMHD equations(\ref{eq:eq1})-(\ref{eq:eq3}).
Magnetic reconnection develops even for small values of the resistivity, but in this section we 
are interested in the diffusive effects on the linear dynamics,
i.e.\   when nonlinear terms are negligible or artificially suppressed 
(we will discuss  such a case in our Conclusions). 
We now consider the effect of standard diffusion 
[case $n=1$ in eqs.~(\ref{eq:eq1})-(\ref{eq:eq2})] on the solutions
(\ref{eq:lin1})-(\ref{eq:lin2b}): these are the solutions of the linearized
equations obtained from (\ref{eq:eq1})-(\ref{eq:eq2}) retaining
also the diffusive terms.

In the \emph{linear regime}, as the magnetic field grows in time 
[(\ref{eq:lin1}), (\ref{eq:lin1s}), (\ref{eq:lin1b})], the diffusive term 
[$\nabla_{_{\!\perp}}^2\, \mathbf{b_{{_\perp}}} \! \propto \! \mathbf{b_{_\perp}}/\ell^2$]
becomes increasingly bigger until diffusion balances the magnetic field growth, 
and the system reaches a saturated equilibrium state. 
Considering diffusion, the magnetic field will evolve as
{\setlength\arraycolsep{-20pt}
\begin{eqnarray}
&&\mathbf{b_{_\perp}} (x,y,z,t) = 
\left[ \mathbf{u^L} (x,y) - \mathbf{u^0} (x,y) \right] 
\nonumber \\
&&\qquad \qquad \qquad \qquad \qquad
\times \, \frac{\tau_R}{\tau_A} \, 
\left[ 1 - \exp \left( - \frac{t}{\tau_R} \right) \right].
\label{eq:diff1}
\end{eqnarray}
}The diffusive timescale associated with the Reynolds number
$Re$ is  $\tau_R = \ell_c^2\, Re / (2\pi)^2$ where $\ell_c$ is the
length-scale of the forcing pattern, that for the pattern (\ref{eq:f0})-(\ref{eq:f1})
is given by $\ell_c = \ell / 4$ where $\ell$ is the orthogonal computational box length.

The total magnetic energy $E_M$ and ohmic dissipation rate $J$ will then be given
by
{\setlength\arraycolsep{-20pt}
\begin{eqnarray}
&&E_M = \frac{1}{2}\, \int_V\! \mathrm{d}^{^3}\hspace{-.4em} x \ \mathbf{b_{_\perp}}^2 = 
E_M^{sat} \, \left[ 1 - \exp \left( - \frac{t}{\tau_R} \right) \right]^2,
\label{eq:diff2} \\
&&J = \frac{1}{R}\, \int_V\! \mathrm{d}^{^3}\hspace{-.4em}  x \ \mathbf{j}^2 = 
J^{sat} \, \left[ 1 - \exp \left( - \frac{t}{\tau_R} \right) \right]^2,
\label{eq:diff3}
\end{eqnarray}
}where $E_M^{sat}$ and $J^{sat}$ are the saturations value reached for
$t \gtrsim 2\, \tau_R$, whose values are given by:
\begin{equation}
E_M^{sat} = \frac{\ell^6 c_A^2 u_{ph}^2 Re^2}{2L(8\pi)^4}, \qquad
J^{sat}  = \frac{\ell^4 c_A^2 u_{ph}^2 Re}{L(8\pi)^2}.
\label{eq:diff4}
\end{equation}
Magnetic energy saturates to a value proportional to the square of both the Reynolds number
and the Alfv\'en velocity, while the heating rate saturates to a value that is proportional to the 
Reynolds number and the square of the axial Alfv\'en velocity.
In general the equations of RMHD are valid as far as the orthogonal magnetic field 
$\mathbf{b_{_\perp}}$ is small compared to the dominant axial field
$\mathbf{B_0} = B_0\, \mathbf{\hat{e}_z}$. In particular incompressibility
holds as far as the perturbed magnetic pressure can be neglected
compared to that of the strong field $\mathbf{b^2_{_\perp}} \ll \mathbf{c^2_A}$.
Therefore the solutions found in this section are valid as far as the saturated
values of the magnetic field satisfy the previous condition. In all the
simulations presented here this condition is satisfied.

\section{Effective Diffusivity: One-point Closure Models} \label{sec:ed}

Given the complexity of the Parker problem, simplified models have been derived.
\cite{hp92} have developed an \emph{effective diffusivity} model that is in effect
a \emph{one-point closure} model. In order to discuss the impact of our work on
their results, we briefly summarize the relevant one-point closure theory \citep{bis03}.

In order to investigate basic properties of the Parker model and compare them with observational
constraints, such as the global heating rate and required energy flux,
the detailed dynamics of turbulent fluctuations (that are essential to determine how the individual
field-lines are heated and hence how radiation is emitted) contain more informations 
than actually needed.

It may then be attempted to split the velocity and magnetic fields into average and fluctuating parts:
\begin{equation}
\mathbf{B} =\, \langle \mathbf{B} \rangle +\, \mathbf{\widetilde{b}}, \qquad
\mathbf{u} =\,  \langle \mathbf{u} \rangle +\, \mathbf{\widetilde{u}}.
\end{equation}
Incompressible MHD equations give for mean fields the following equations:
{\setlength\arraycolsep{-10pt}
\begin{eqnarray}
&&\partial_t \langle \mathbf{u} \rangle  + 
 \langle \mathbf{u} \rangle \cdot \nabla \langle \mathbf{u} \rangle = 
- \nabla \langle P\rangle + \langle \mathbf{B} \rangle \cdot \nabla \langle \mathbf{B} \rangle
\nonumber \\[.1em]
&& \qquad \qquad \qquad \qquad \quad\
- \nabla \cdot \langle \mathbf{\widetilde{u}} \mathbf{\widetilde{u}} -  \mathbf{\widetilde{b}} \mathbf{\widetilde{b}} \rangle
+ \nu\, \nabla^2 \langle \mathbf{u} \rangle,
\label{eq:aeq1} \\[.5em]
&&\partial_t \langle \mathbf{B} \rangle  + 
\langle \mathbf{u} \rangle \cdot \nabla \langle \mathbf{B} \rangle = 
\langle  \mathbf{B} \rangle \cdot \nabla \langle \mathbf{u} \rangle 
\nonumber \\[.1em]
&& \qquad \qquad \qquad \qquad \qquad 
+ \nabla \times \langle \mathbf{\widetilde{u}} \times \mathbf{\widetilde{b}} \rangle
+ \eta\, \nabla^2\, \langle \mathbf{B} \rangle,
\label{eq:aeq2} \\[.5em]
&& \nabla \cdot \langle \mathbf{u} \rangle = \nabla \cdot \mathbf{\widetilde{u}} =
\nabla \cdot \langle \mathbf{B} \rangle = \nabla \cdot \mathbf{\widetilde{b}} = 0,
\label{eq:aeq3} 
\end{eqnarray}
}where symbols have the usual meaning, and in particular $\nu$ and $\eta$ are the microscopic viscosity
and resistivity of the plasma. If it is possible to model the terms that contain the small-scale fluctuations
then eqs.~(\ref{eq:aeq1})-(\ref{eq:aeq3}) allow to advance the mean fields.

\begin{table}
\begin{center}
\caption{Summary of  the simulations\label{tbl}}
\begin{tabular}{ccllcc}
\hline \hline\\[-5pt]
Run & $c_A$ & $n_x \times n_y \times n_z$ & forcing & $n$ & $Re$ or $Re_4$ \\[5pt]
\hline\\[-5pt]
A........ & 200 & 512 x 512 x 200 & \emph{shear}: t    &1 & 800 \\[2pt]
B........ & 200 & 256 x 256 x 100 & \emph{shear}: t    &1 & 400 \\[2pt]
C........ & 200 & 128 x 128 x 50   & \emph{shear}: t    &1 & 200 \\[2pt]
D........ & 200 & 128 x 128 x 50   & \emph{shear}: t    &1 & 100 \\[2pt]
E........ & 200 & 128 x 128 x 50   & \emph{shear}: t    &1 &   10 \\[2pt]
F........ & 200 & 512 x 512 x 200 & \emph{shear}: t,b &4 & $10^{19}$ \\[2pt]
G........ & 200 & 512 x 512 x 200 & \emph{vortex}: t,b &4 & $10^{19}$ \\[5pt]
\hline
\end{tabular}
\tablecomments{$c_A$ is the axial Alfv\'en velocity, $n_x \times n_y \times n_z$
is the numerical grid resolution, b or t in the forcing column indicates whether the 
forcing is applied at the bottom and/or top plates. n is the \emph{dissipativity},
$n=1$ indicates normal diffusion, $n=4$ hyperdiffusion. The next column indicates
respectively the value of the Reynolds number $Re$ or of the hyperdiffusion 
coefficient $Re_4$.}
\end{center}
\end{table}

This is a two-scale approach, and the average and fluctuating parts can be represented 
as the large-scale and small-scale fields or, introducing a suitable cut-off wavenumber $K$
(for instance the grid resolution if this is applied to numerical simulations),
as the \emph{low-} and \emph{high-pass filtered}  ``lesser'' and ``greater'' functions
\begin{eqnarray}
\langle \mathbf{B} \rangle & = & \mathbf{B}^{<}_{K} = 
\sum_{k \le K} \mathbf{\widehat{B}_{k}}\, e^{i \mathbf{k} \cdot \mathbf{x}},\\
\mathbf{\widetilde{b}} & = & \mathbf{B}^{>}_{K} = 
\sum_{k > K} \mathbf{\widehat{B}_{k}}\, e^{i \mathbf{k} \cdot \mathbf{x}}.
\end{eqnarray}
The relevant quantities to be modeled  in eqs.~(\ref{eq:aeq1})-(\ref{eq:aeq3}) are the 
``turbulent'' stress tensors
\begin{equation}
R_{ij} = - \langle \widetilde{u}_i \widetilde{u}_j - \widetilde{b}_i \widetilde{b}_j \rangle, \qquad
S_{ij} = - \langle \widetilde{u}_i \widetilde{b}_j - \widetilde{u}_j \widetilde{b}_i \rangle.
\end{equation}
Phenomenological modeling and numerical simulations \citep{yo90,yo91} have shown that these stress
tensor can be approximated with
\begin{eqnarray}
&&R_{ij} \sim \nu_t  \Big(   \partial_i \langle  u_j \rangle +  \partial_j \langle u_i  \rangle \Big), \label{eq:tdiff1} \\
&&S_{ij} \sim \eta_t \Big(   \partial_i \langle  B_j \rangle -  \partial_j \langle B_i  \rangle \Big), \label{eq:tdiff2}
\end{eqnarray}
where the coefficients $\nu_t$ and $\eta_t$ are called turbulent viscosity and resistivity, and for a plasma
in coronal conditions have much higher values than $\nu$ and $\eta$.
Inserting (\ref{eq:tdiff1})-(\ref{eq:tdiff2}) in the equations for the mean fields (\ref{eq:aeq1})-(\ref{eq:aeq2}),
we obtain a set of equations for the mean fields that has \emph{the same structure of incompressible MHD
equations} except that $\nu \rightarrow \nu_t$ and $\eta \rightarrow \eta_t$.

Then for a system embedded in a strong axial field \citep{mon82} equations~(\ref{eq:eq1})-(\ref{eq:eq3}) 
that we use to model the system are also  a one-point closure model if the dissipative coefficients are 
considered as \emph{effective} ones.

\cite{hp92} use the results of an eddy damped quasi-normal Markovian approximation \citep{pfl76}
to express  the \emph{effective diffusivity} coefficients  $\nu_t$ and $\eta_t$ as a function 
of the energy flux $\epsilon$ flowing along the inertial range. 

Additionally they \emph{suppose} that the 1D boundary forcing velocity leads 
the large-scales to evolve in a laminar (fields are 1D and directed along $y$, the same 
direction of the forcing) steady state ($\partial_t =0$).
Equations~(\ref{eq:eq1})-(\ref{eq:eq3}) and their boundary forcing 
[of which our forcing (\ref{eq:f0})-(\ref{eq:f1}) is representative] are used to compute the flux 
of energy $S$ entering the system due to the dragging of the field-lines footpoints.  

$S$ is a function of the system parameters and the effective 
diffusivity coefficients $\nu_t$ and $\eta_t$ that are the only unknown variables in both energy fluxes 
$S$ and $\epsilon$.
The solution of the problem is achieved requiring that the flux $S$ entering the system at the large scales
is equal to the flux $\epsilon$ flowing from the large to the small scales.

As a matter of fact  their \emph{laminar steady state coincides with the saturation linear state} that we have computed
in the previous section (\S~\ref{par3}). This was obtained neglecting the nonlinear terms in 
eqs.~(\ref{eq:eq1})-(\ref{eq:eq2}) and retaining the diffusive terms. 
They obtain it in a similar way, by \emph{supposing} that the induced fields
will retain the 1D symmetry of the forcing. Thus the nonlinear terms vanish, as can be proved by
direct substitutions of the generic fields $\mathbf{b_{_\perp}} = f(x)\, \mathbf{\hat{e}_y}$,  
$\mathbf{u_{_\perp}} = g(x)\, \mathbf{\hat{e}_y}$, with $f$ and $g$ generic functions.

Our simulations investigate the large-scale dynamics and 
as we show in \S~\ref{sec:ns} the large-scale flow is not laminar and it is steady 
only in a statistical sense.
Turbulence cannot be confined only to the small scales. We discuss the implications for the findings
and scaling laws of \cite{hp92} in section \S~\ref{sec:con} devoted to our conclusions.

\section{Numerical Simulations} \label{sec:ns}
 
In this section we present a series of numerical simulations, summarized in 
Table~\ref{tbl}. We will first describe the results of simulations A--E
that model with different resolutions and associated different Reynolds numbers
a coronal layer driven by the sheared velocity pattern~(\ref{eq:f0})
at the top plate $z=L$ and a vanishing velocity~(\ref{eq:f1}) at the bottom plate 
$z=0$. In all simulations the computational box has an aspect ratio of $10$
with $\ell = 1$ and $L=10$.

\subsection{Shear Forcing: run~A} \label{sec:runa}

We present here the results of run~A, a simulation performed with a numerical grid
of $512 \times 512 \times 200$ points, normal (n=1) diffusion with a Reynolds number
$Re=800$. The Alfv\'en velocity is $v_A = 200\, km\, s^{-1}$ corresponding to a ratio
$c_A = v_A/u_{ph} = 200$. The total duration is  $600$ axial Alfv\'en crossing
times $\tau_A = L/v_A$. 

\begin{figure}
\plotone{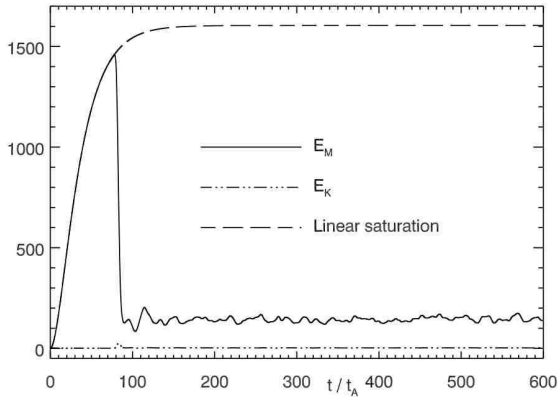}
\caption{\emph{Run~A:} Magnetic ($E_M$) and kinetic ($E_K$) energies
as a function of time ($\tau_A = L/v_A$ is the axial Alfv\'en crossing time).
The dashed curve shows the time evolution of magnetic energy
if the system were unperturbed [eq.~(\ref{eq:diff2})].
\label{fig1}}
\end{figure}

\begin{figure}
\plotone{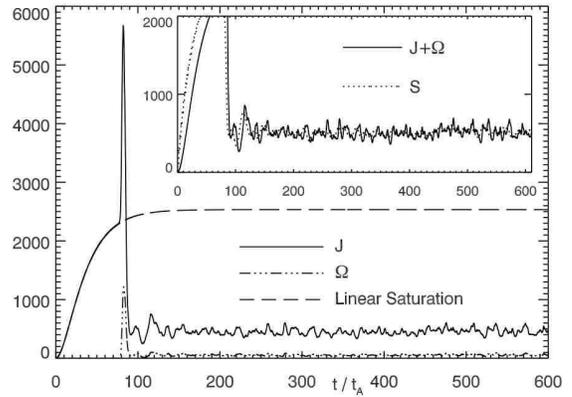}
\caption{\emph{Run~A:} Ohmic ($J$) and viscous ($\Omega$) dissipation rates as a 
function of time. The dashed curve is the linear saturation curve for
the ohmic dissipation rate  [eq.~(\ref{eq:diff3})]. In the inset the integrated 
Poynting flux $S$ is shown to dynamically balance the total dissipation.
\label{fig2}}
\end{figure}

We add to the system a perturbation (naturally present in the coronal environment)
 for the magnetic and velocity fields inside the
computational box at time $t=0$. If there were \emph{no perturbation} the system
would follow the \emph{linear saturation curves} plotted with dashed lines
in Figures~\ref{fig1} and \ref{fig2}, as discussed in \S\ref{par3}.

We have not performed a full linear instability analysis, but we
have used as perturbations either an Orszag-Tang vortex\citep{ot79}, i.e.
\begin{eqnarray}
&& \delta \mathbf{u} = \epsilon\, \bigg[ - \sin \left( 2\pi y \right) \mathbf{\hat{e}_x}
+  \sin \left( 2\pi x \right) \mathbf{\hat{e}_y} \bigg] \label{eq:ot1}\\[5pt]
&& \delta \mathbf{b} = \epsilon\, \bigg[ - \sin \left( 4\pi y \right) \mathbf{\hat{e}_x}
+  \sin \left( 2\pi x \right) \mathbf{\hat{e}_y} \bigg] \label{eq:ot2}
\end{eqnarray}
or a ``white noise'' (i.e.\ the value of the potentials associated with the fields are given
in each grid point a random number included between 0 and 1),
with different amplitudes $\epsilon$.
We have computed the resulting growth rates $\gamma$ with our nonlinear code.

For both perturbations the value of the growth rates is high and similar to each other.
The white noise has a slightly higher growth rate at  $\gamma \tau_A \sim 0.69$.
This timescale is of the order of the Alfv\'en crossing time $\tau_A$, which corresponds
to an ideal timescale.
This implies that the Orszag-Tang vortex is close to the most unstable
mode, which is selected from all the modes that are excited by the random
perturbation. As we shall see in the following the instability is a multiple tearing
mode.

For each kind of perturbation,
the smaller the amplitude the later the system becomes unstable.
For the simulation presented in this section we have used a ``white noise''
perturbation with an amplitude $\epsilon =  10^{-16}$, very small compared 
to the boundary imposed velocity eq.~(\ref{eq:f0}) which is $\mathcal{O}(1)$.

In Figures~\ref{fig1}-\ref{fig2} we plot  the total magnetic and kinetic energies
\begin{equation}
E_M = \frac{1}{2} \int \! \mathrm{d}V\, \mathbf{b_{_\perp}}^2, \qquad
E_K = \frac{1}{2} \int \! \mathrm{d}V\, \mathbf{u_{_\perp}}^2, 
\end{equation}
and the total ohmic and viscous dissipation rates
\begin{equation}
J             = \frac{1}{Re} \int \! \mathrm{d}V\, \mathbf{j}^2, \qquad
\Omega = \frac{1}{Re} \int \! \mathrm{d}V\, \mathbf{\omega}^2, 
\end{equation}
along with the saturation  curves~(\ref{eq:diff2})-(\ref{eq:diff3}).
For smaller values of the perturbation amplitude $\epsilon$ the system
becomes unstable sooner but the ensuing dynamics in the fully nonlinear
stage are similar with same average values for all relevant physical
quantities.

\begin{figure*}
\begin{centering}
\includegraphics[scale=.50]{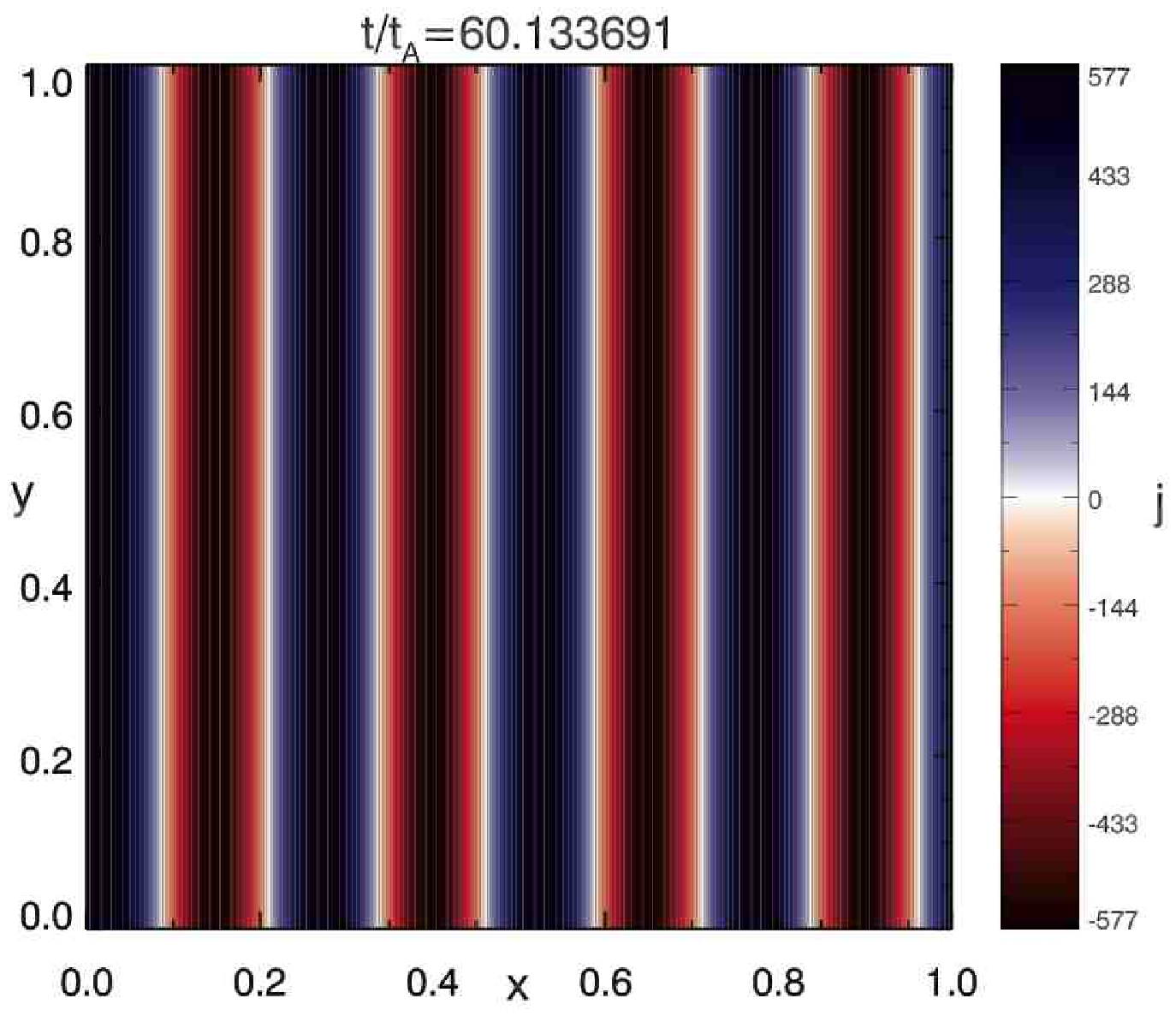}
\includegraphics[scale=.50]{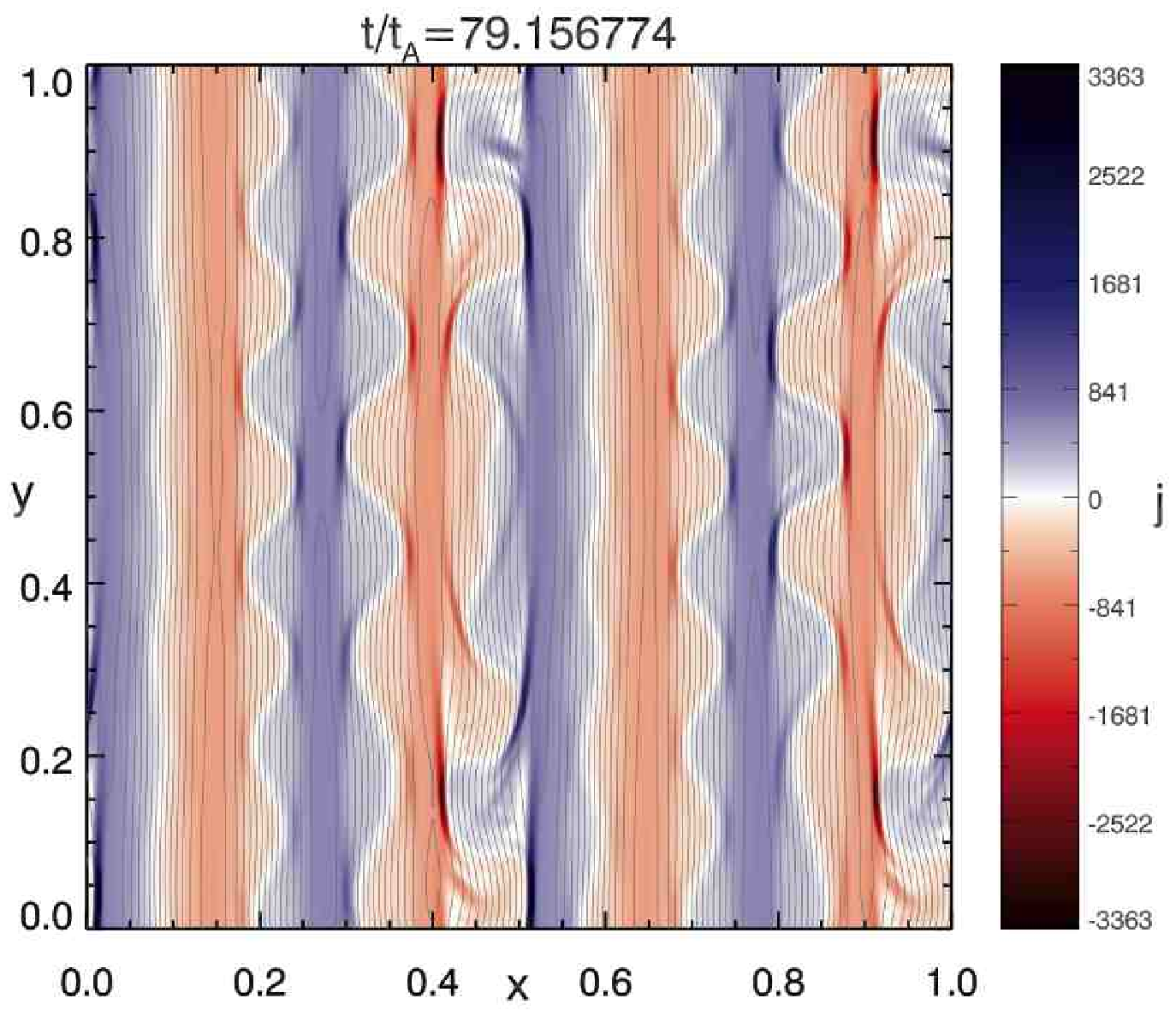}\\[.3em]
\includegraphics[scale=.50]{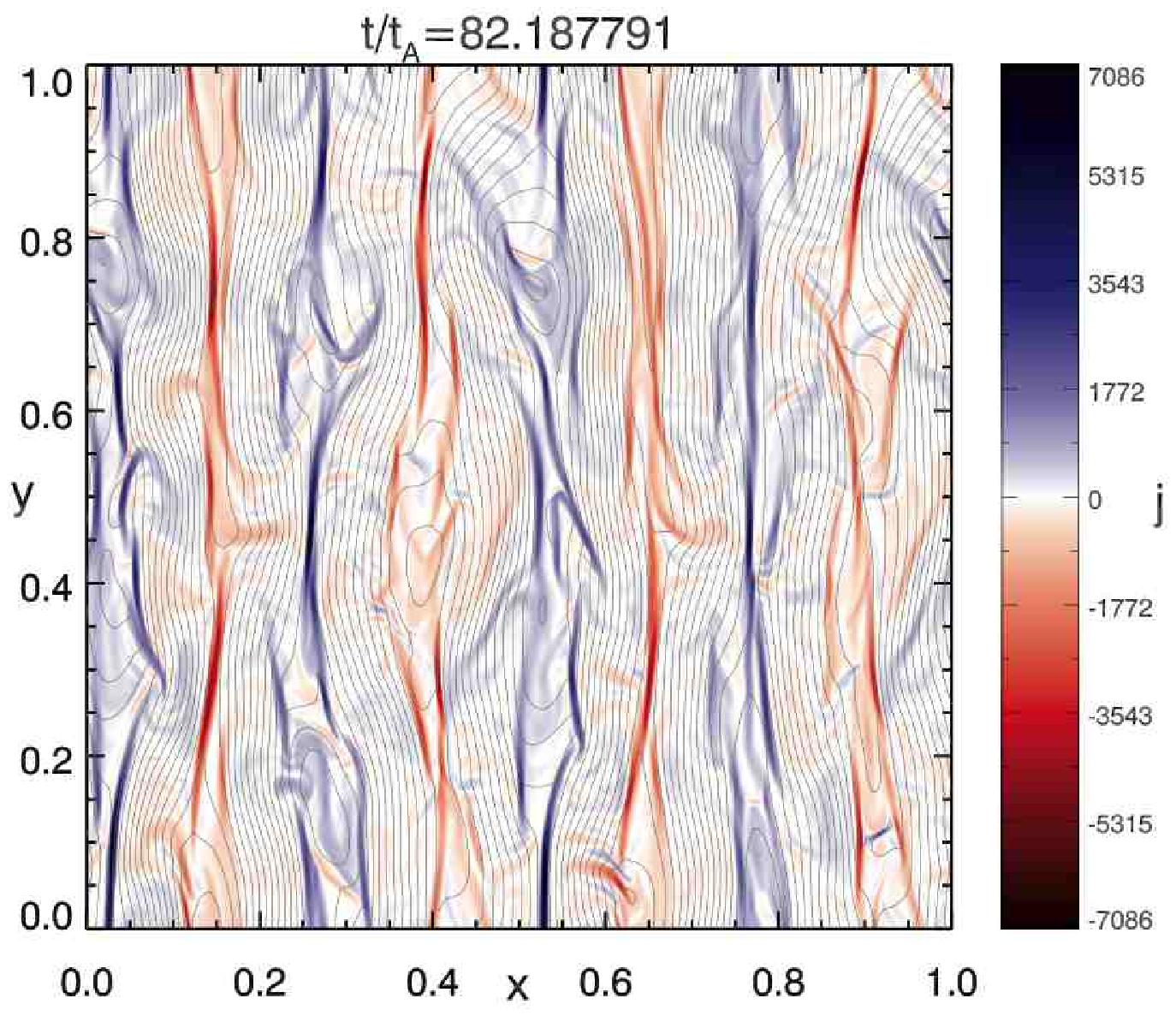}
\includegraphics[scale=.50]{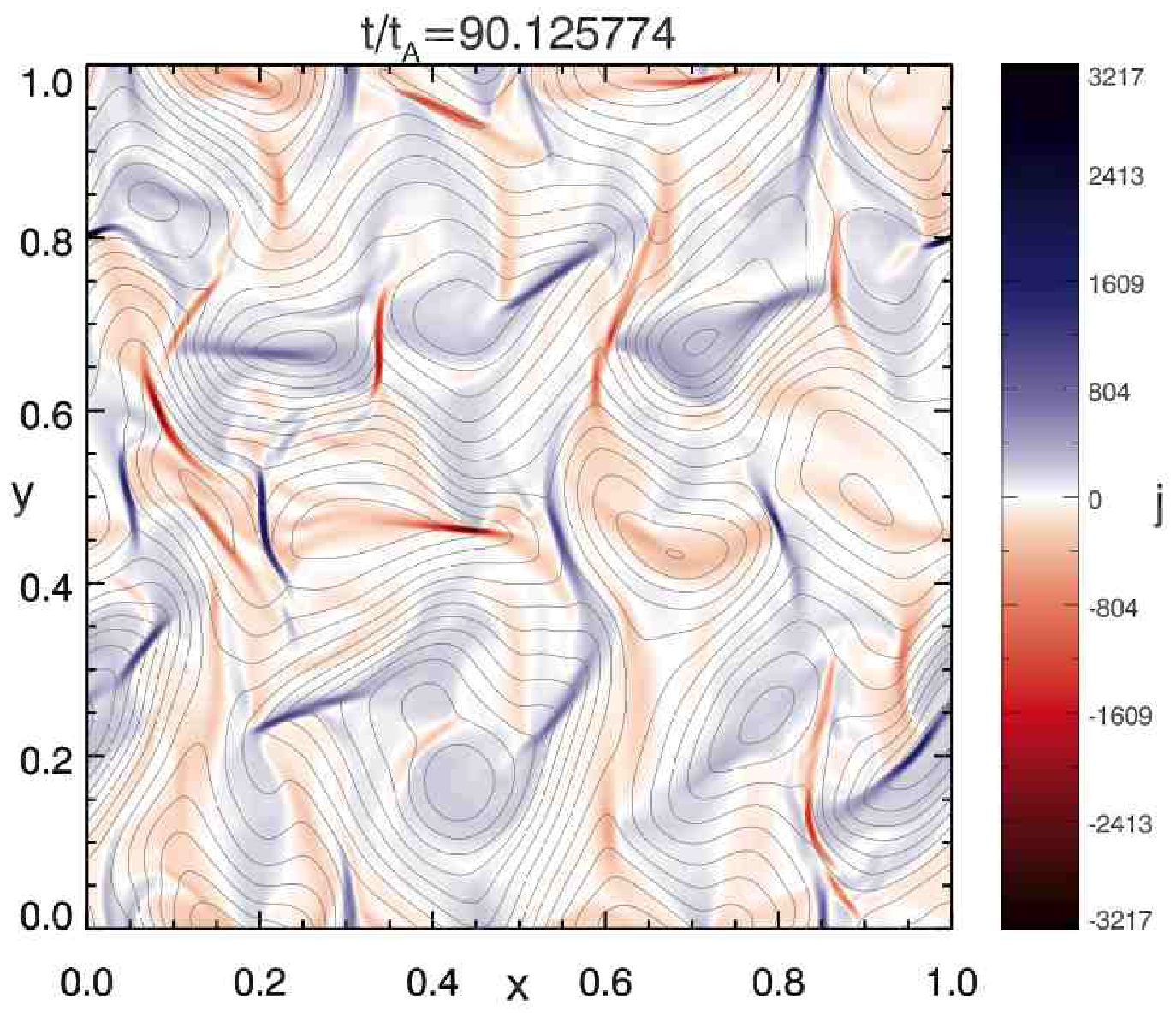}\\[.3em]
\includegraphics[scale=.50]{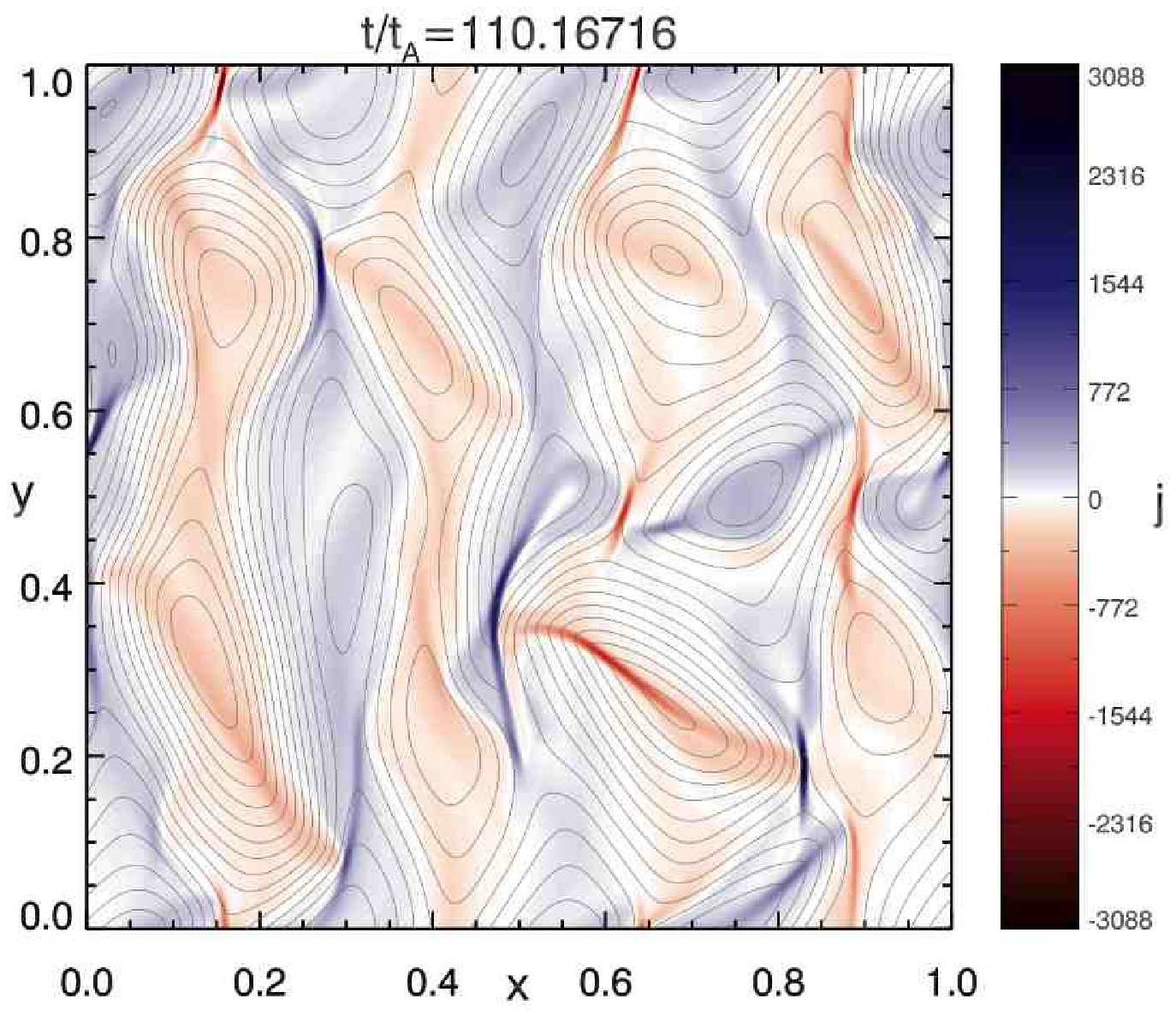}
\includegraphics[scale=.50]{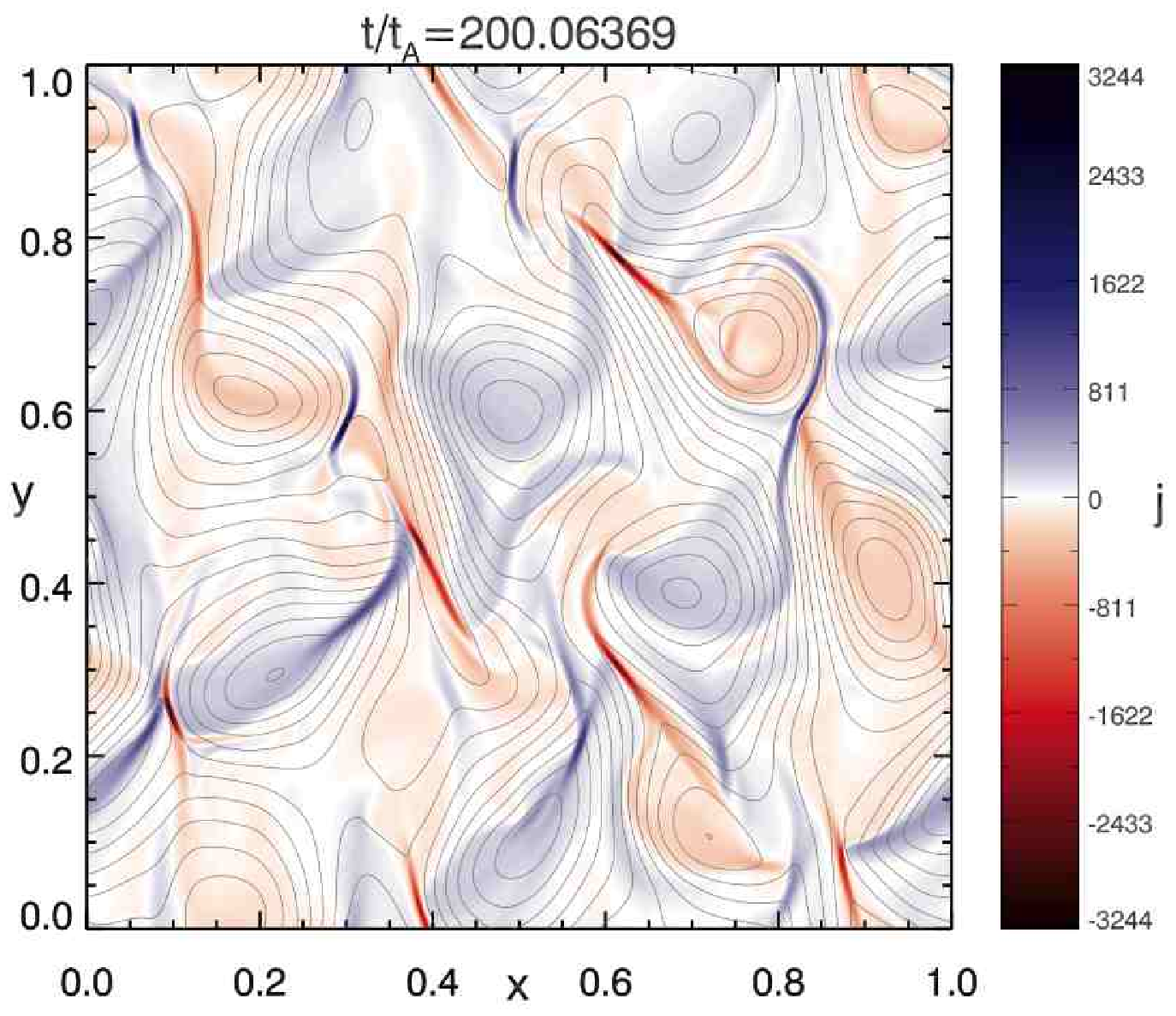}
\caption{\emph{Run~A}: Axial component of the current $j$ (in color) and field-lines 
of the orthogonal magnetic field in the midplane ($z=5$) at selected times covering 
the linear and nonlinear regimes up to $t\sim 200\, \tau_A$.
During the linear stage ($t \sim 60\, \tau_A$) the orthogonal magnetic field 
is a mapping of the boundary shear velocity [eq.~(\ref{eq:f0})]. After the transition to 
the nonlinear stage due to a multiple tearing instability ($t \sim 79$ and $82\, \tau_A$)
the topology of the magnetic field departs from the boundary velocity mapping and 
evolves dynamically in time ($t \sim 90, 110$ and $200\, \tau_A$).
\label{fig3}}
\end{centering}
\end{figure*}

\begin{figure*}
\begin{centering}
\includegraphics[scale=.4]{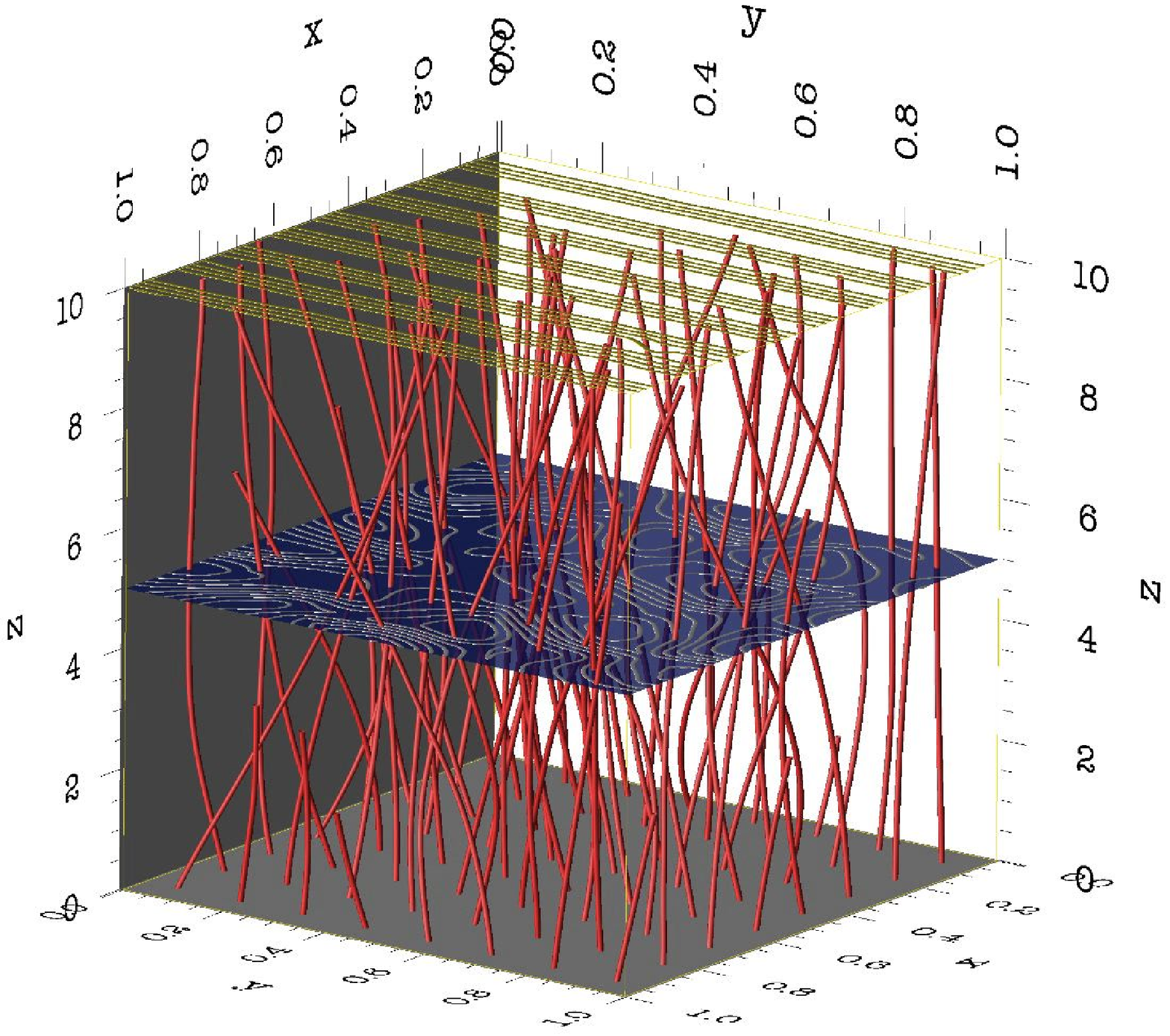}
\includegraphics[scale=.4]{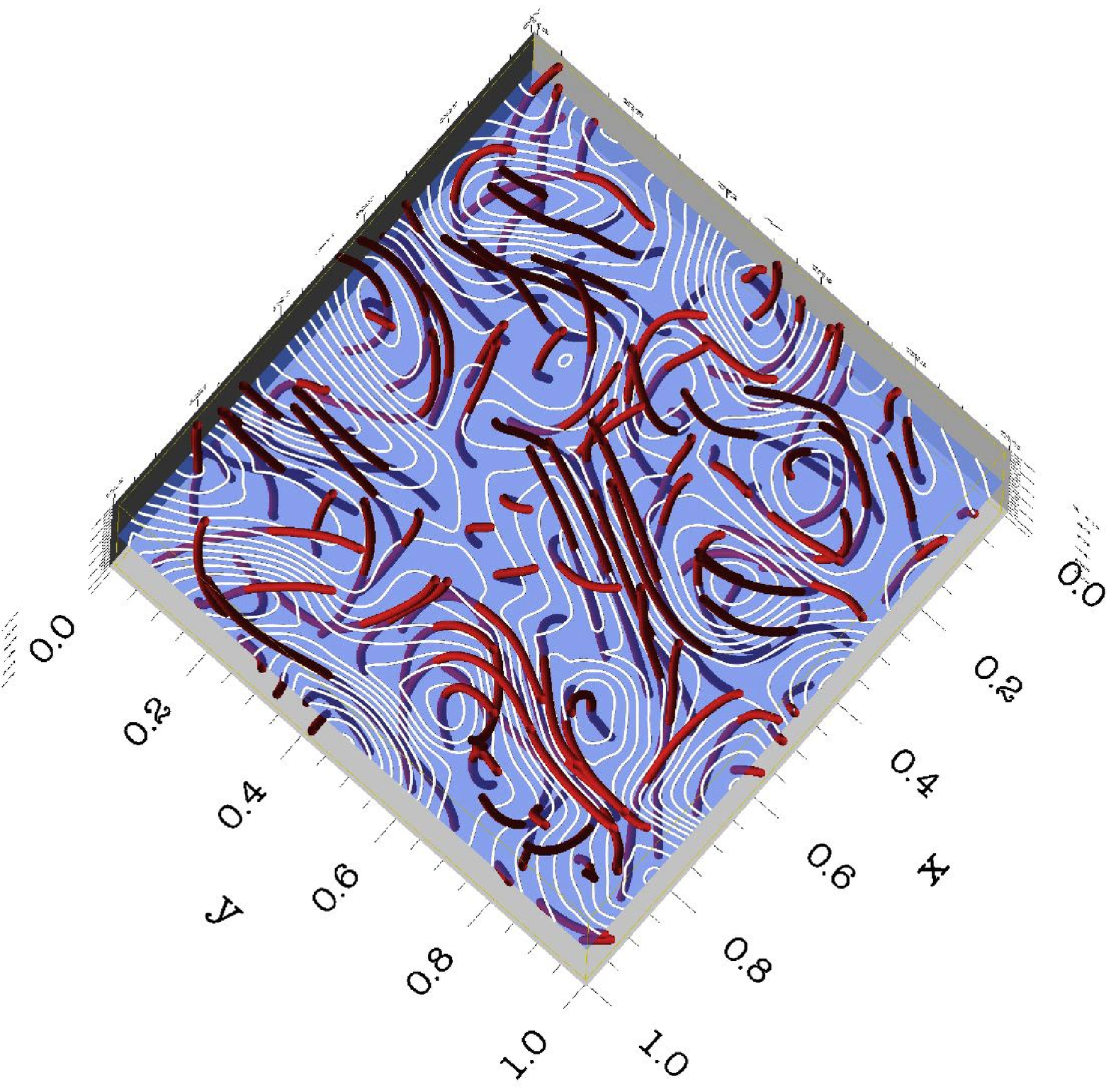}\\[.3em]
\includegraphics[scale=.4]{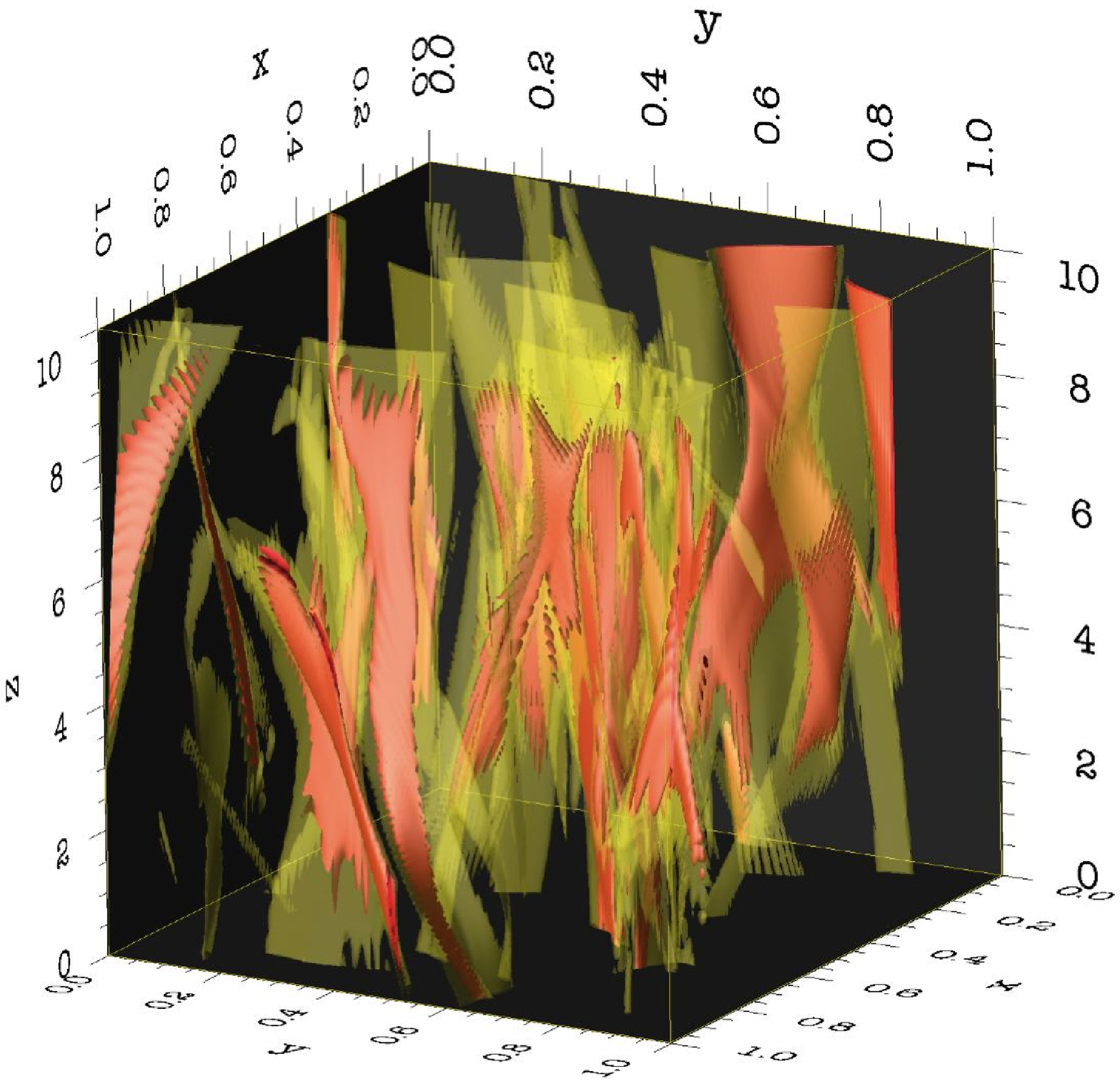}
\includegraphics[scale=.4]{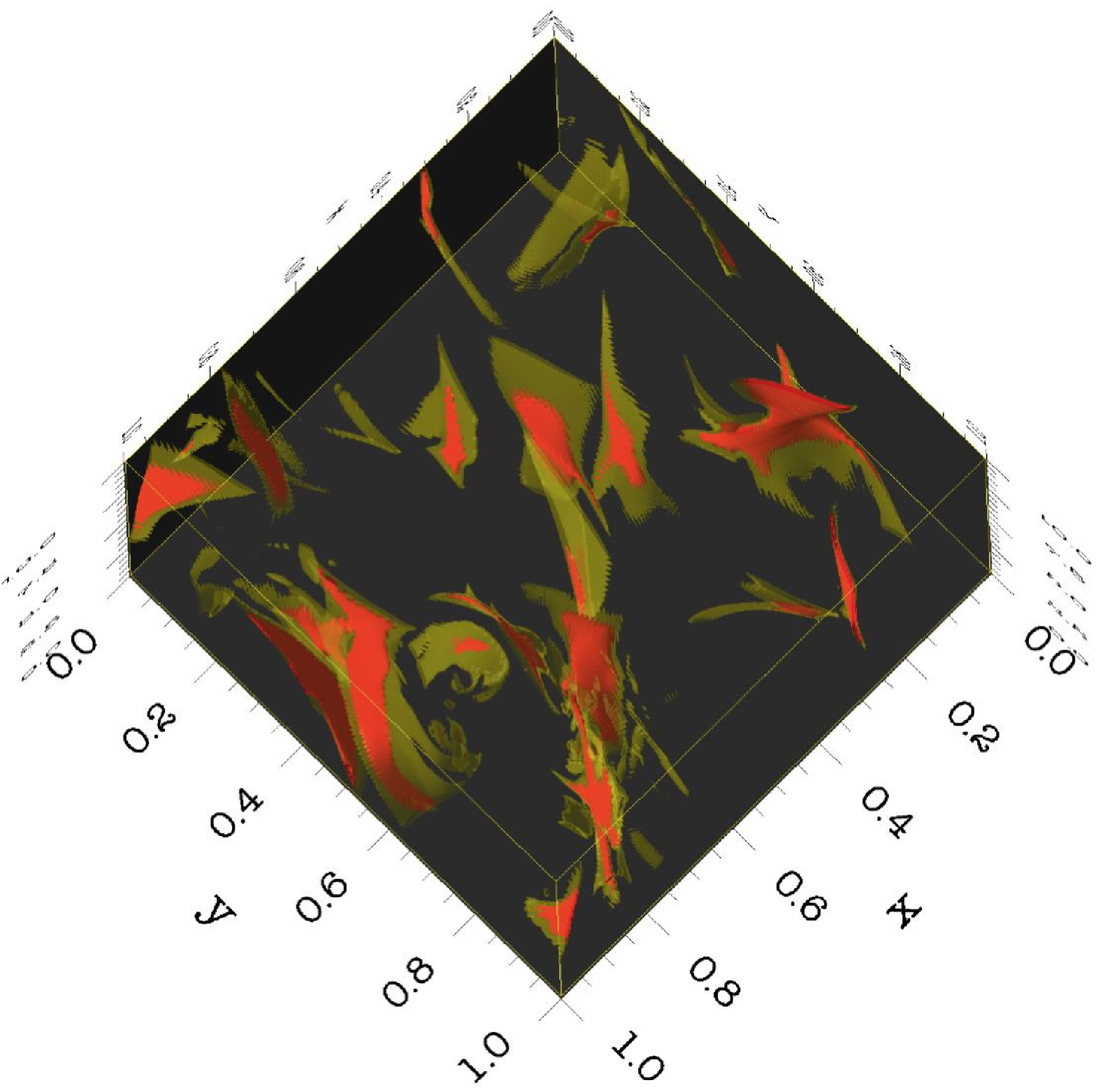}\\[.3em]
\caption{Side and top views of a snapshot of magnetic field-lines 
(\emph{top row}) and current sheets (\emph{bottom row}) at time $\tau \sim 550\, \tau_A$. 
The box has been rescaled for an improved visualization. 
\emph{Top:} Field lines of the total magnetic field (orthogonal 
plus axial), and in the midplane ($z=5$) field-lines of the orthogonal component of the magnetic field.
In the side view we have superimposed in yellow the streamlines of the boundary forcing velocity
in the top plate ($z=10$), at the bottom we impose a vanishing velocity.
\emph{Bottom:} Two isosurfaces of the squared current $j^2$. The isosurface at the value 
$j^2 = 2.8 \times 10^5$ is represented in partially transparent yellow, while red displays 
the isosurface with $j^2 = 8 \times 10^5$, well below the maximum value of the current 
at this time $j^2_{max} = 3.6 \times 10^7$. As is typical of current sheets, isosurfaces 
corresponding to higher values of $j^2$ are nested inside those corresponding to 
lower values. The current sheets filling factor is small.
\label{fig4}}
\end{centering}
\end{figure*}

As shown in Figures~\ref{fig1}-\ref{fig2} and \ref{fig3} the shear velocity at the top
boundary ($z=10$) initially induces velocity and magnetic fields inside the 
volume that follow the linear behavior  
[eq.~(\ref{eq:lin2}) and (\ref{eq:diff1})].
The magnetic energy and the ohmic dissipation rate also follow initially
the linear diffusive saturation curves eq.~(\ref{eq:diff2})-(\ref{eq:diff3}). From eq.~(\ref{eq:diff4})
with the values for this simulation we have that the magnetic energy
and ohmic dissipation would reach the saturation values
\begin{equation}
E_M^{sat} = 1604, \qquad J^{sat} =  2533,
\end{equation}
with a diffusion time $\tau_R \sim 25\, \tau_A$, if the system were unperturbed.

\begin{figure*}
\includegraphics[scale=.4]{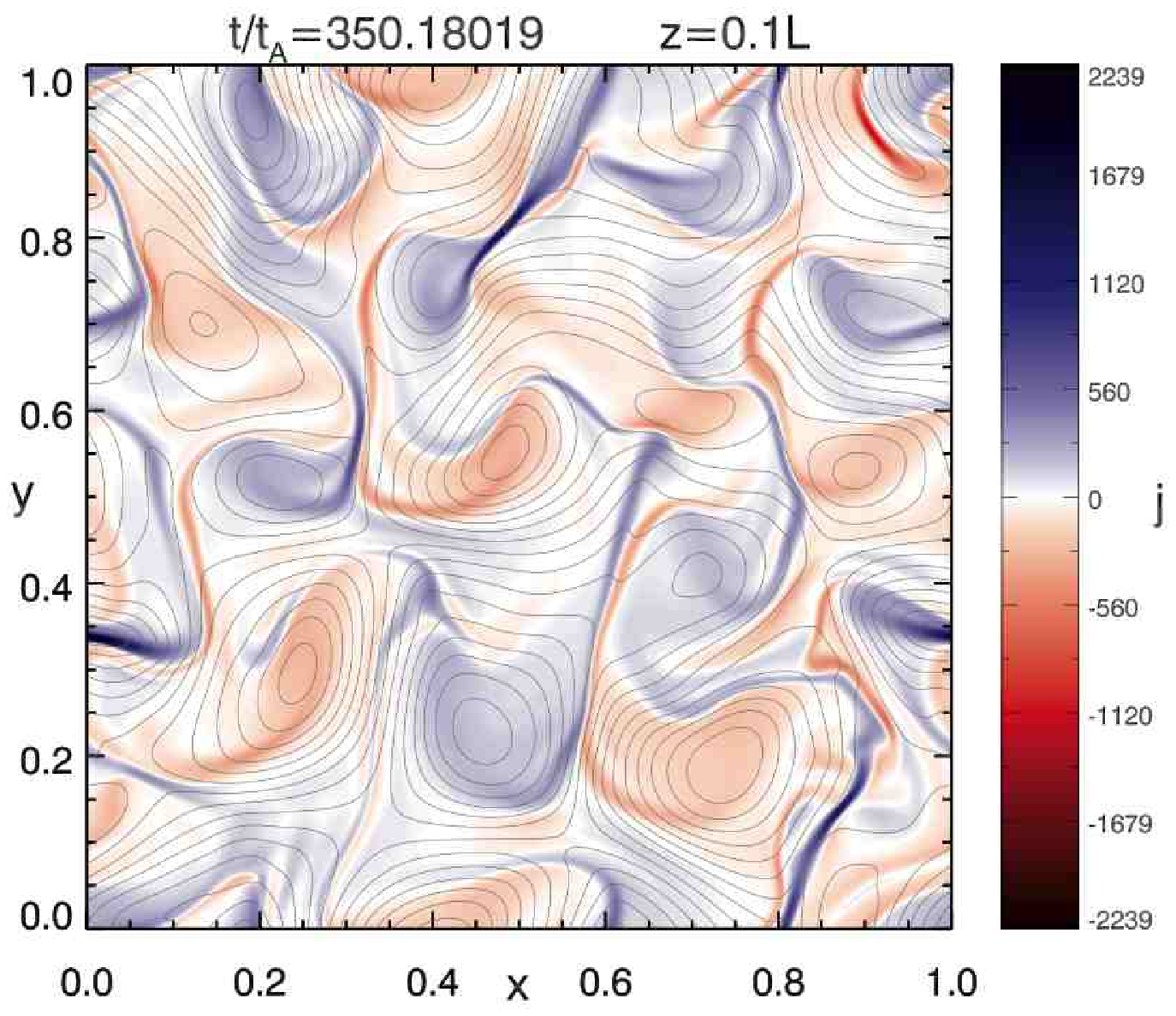}
\includegraphics[scale=.4]{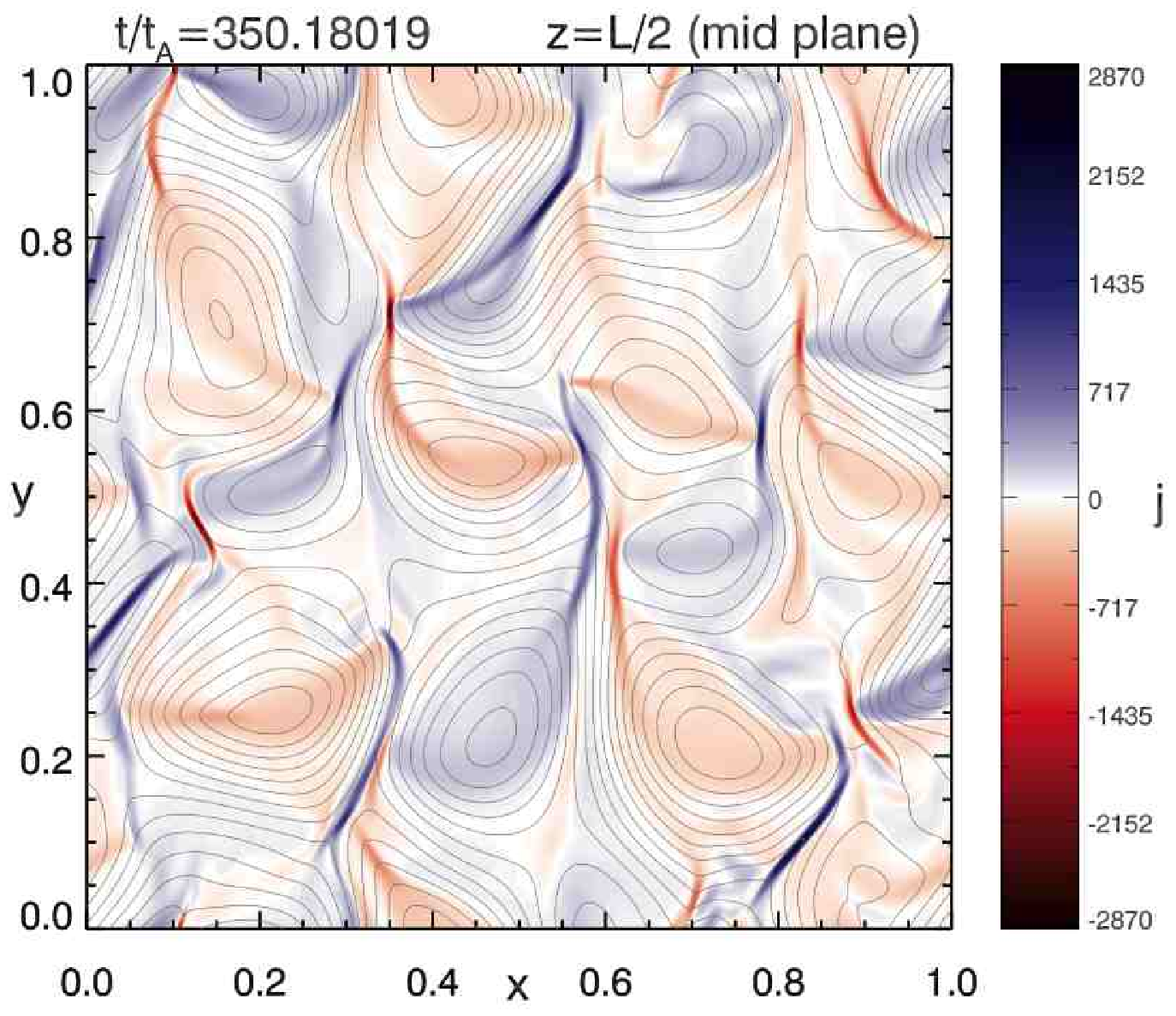}
\includegraphics[scale=.4]{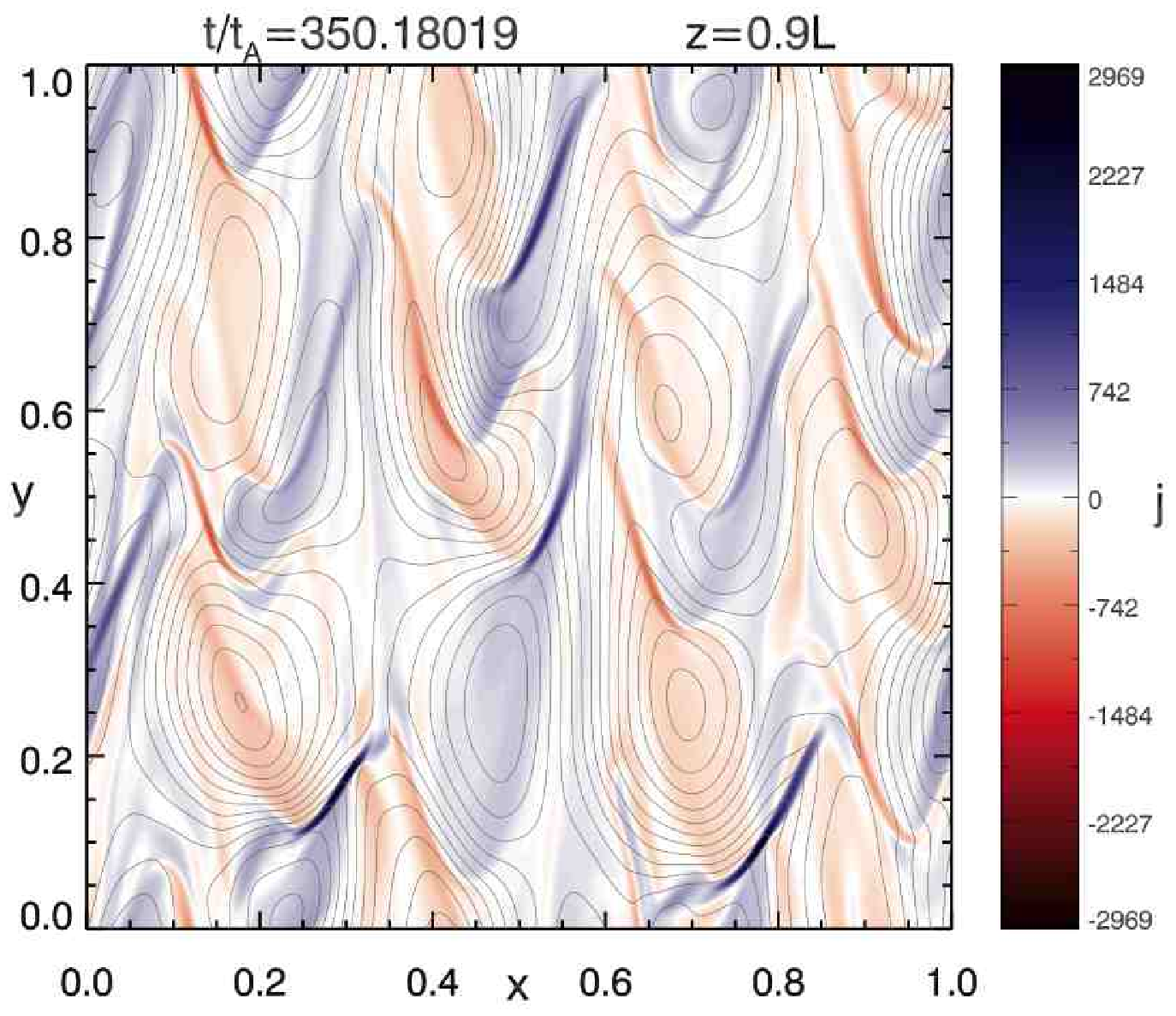}
\caption{\emph{Run~A}: Snapshots at time $t = 350\, \tau_A$ of the axial component of 
the current $j$ (in color) and field-lines of the orthogonal magnetic field in 3 distinct x-y planes:
the plane $z=1$ close to the bottom unforced boundary, the midplane $z=5$, and in plane 
$z=9$ close to the forced boundary $z=10$. The box lenght is $L=10$.
\label{fig5}}
\end{figure*}

Figure~\ref{fig3} shows  snapshots of the magnetic field-lines 
(orthogonal component $\mathbf{b_{_\perp}}$) and
electric current at selected times in the mid-plane $z=5$.
Until time $t \sim 75\, \tau_A$ the velocity and magnetic fields in 
the volume are a mapping of the velocity at the boundary 
[eq.~(\ref{eq:lin2}) and (\ref{eq:diff1})], i.e.\
the sheared velocity at the boundary induces a sheared magnetic field in the volume.

This magnetic configuration is well known to be unstable to tearing instabilities
\citep{fkr63}, 
and in fact around time $t \sim 79\, \tau_A$ a multiple tearing instability develops 
when the system transitions from the linear  to the nonlinear stage. 
Magnetic field-lines reconnect in correspondence of X-points while the characteristic
magnetic islands are formed. The randomness of the perturbations is reflected in
the lack of symmetry of the reconnecting field.

The magnetic energy (Figure~\ref{fig1})
accumulated up to this point is then released in a big burst of ohmic dissipation 
(see Figure~\ref{fig2}). Notice that correspondingly 
there is a peak in the viscous dissipation (i.e.\ the integrated vorticity), but it is smaller
that the electric current peak as in the following dynamical evolution.
This is in fact a magnetically dominated system also when considering only
the magnetic fluctuations $\mathbf{b_{_\perp}}$ with respect to velocity fluctuations
$\mathbf{u_{_\perp}}$, without taking into account the big axial field $B_0$.

Up to this point the dynamics are not surprising, the magnetic field gets sheared and
the shear  increases in time. Such a configuration is very well known to be unstable
to reconnection, and then it is expected that a tearing-like instability should develop.

What is surprising are the dynamics at later times. In fact as can be seen in 
Figures~\ref{fig1}-\ref{fig2}  for $t > 90\, \tau_A$ the system reaches a \emph{magnetically
dominated statistically steady state} where integrated quantities fluctuate around an average
value. In particular velocity fluctuations are smaller than magnetic fluctuations,
which in turn are small compared to the axial magnetic field 
($\langle \mathbf{b}_{_\perp}^2 \rangle^{1/2}/B_0 \sim 0.027$). 
Therefore the orthogonal magnetic and velocity fields satisfy
the RMHD ordering that requires them to be small compared to the
axial field $B_0$.

The energy power entering the system at the boundaries as a result of the
work done by photospheric motions on the  footpoints of magnetic field-lines
is given by the integrated Pointing flux
\begin{equation}
S = c_A \! \int\limits_{z=L} \! \! \mathrm{d}a\, \mathbf{b_{_\perp}}  \cdot \mathbf{u_{_\perp}^L},
\end{equation}
where $\mathbf{u_{_\perp}^L}$ is the photospheric forcing velocity~(\ref{eq:f0}).

Dissipation rates and the Poynting flux also fluctuate
around a mean value. In particular, as shown in the inset in Figure~\ref{fig2}, the Poynting
flux and the total dissipation (ohmic plus viscous) balance each other on the average, 
although on shorter timescales a lag between the two signals is present. In this steady state the energy
that is injected into the system is then, on the average, completely dissipated.

The surprising feature is that the shearing in the magnetic field inside the volume
is not recreated, as shown in Figure~\ref{fig3}. 
It is natural to think that after the first big dissipative event
around $t \sim 82\, \tau_A$, the shear forcing velocity at the boundary would recreate
over time a sheared magnetic field in the system that should then lead to another
big dissipative event and so on.

The dynamics of this system are in fact commonly approximated as a sequence of
equilibria, each destabilized by magnetic reconnection. 
This approximation is attained by neglecting the velocity and kinetic
pressure in the MHD equations whose solution is then bound to be a static force-free
equilibrium.
Our simulations confirm that the system is magnetically dominated,
and in particular magnetic energy is bigger than kinetic energy, as shown in 
Figure~\ref{fig1} where on the average $E_M \sim 61\, E_k$. But the
self-consistent evolution of the kinetic pressure and velocity, although
small compared with the dominant axial magnetic field $B_0$, does not
bind the system to force-free equilibria and allows the possible development
of alternative dynamics.

In the following sections we will analyze further aspects of the dynamics and
the spectral properties of the system. But first we illustrate the topology of
the magnetic and velocity fields, to understand why a sheared magnetic
field is not recreated.

\subsubsection{Magnetic Field Topology and the Origin of Turbulence} \label{sec411}

As shown in Figure~\ref{fig3} at time $t \sim 79\, \tau_A$ reconnection starts to develop,
enhancing the ohmic dissipation, that reaches a peak around  $t \sim 82\, \tau_A$.
This big dissipative event burns a large fraction of the magnetic energy previously
accumulated by the system (Figure~\ref{fig1}). \emph{But not all the magnetic energy
is dissipated}, which would otherwise 
bring the system to a configuration similar to the initial condition at $t=0$.

This dissipative event is in fact due to magnetic reconnection, that during
its evolution produces a component of the magnetic field along $x$, 
the cross-shear direction, forming magnetic islands [see Figure~\ref{fig3} at times 
$t \sim 79$ and $82\, \tau_A$].
In fact around $t \sim 90\, \tau_A$, at the end of the big dissipative event,
the topology of the orthogonal component of the magnetic field is characterized
by magnetic islands. Naturally the Lorentz force does not vanish now and
the vorticity is not constant along the streamlines. As typical of magnetic
reconnection vorticity forms misaligned quadrupolar structures around current 
sheets [see \cite{rved08}].

Although the forcing velocity at the boundary is always a shear [eqs.~(\ref{eq:f0})-(\ref{eq:f1})],
\emph{nonlinear terms do not vanish} as they do during the linear stage for $t < 79\, \tau_A$.
When they vanish  magnetic energy can be stored without getting dissipated (see \S\ref{par3}).
But now nonlinearity can \emph{redistribute} along the cross-shear ($x$) direction part of the energy
associated with the shear-aligned ($y$-oriented) field along which the forcing injects energy,
and continuously cascades to lower scales as described in the following sections.

The three-dimensional structures are shown in Figure~\ref{fig4}.
Although the magnetic energy dominates over the kinetic energy, the ratio of the rms of the 
orthogonal magnetic field over the axial dominant field $B_0$ is quite small. 
For $c_A = 200$ it is $\sim 3\%$, so that the average inclination of the magnetic field-lines 
with respect to the axial direction is just $\sim 2^{\circ}$, it is only for lower value
of $c_A$ that this ratio increases and the angle increases accordingly \citep{rved08}.
The field-lines of the total magnetic field at time $550\, \tau_A$ are shown in 
Figure~\ref{fig4} (\emph{top row}). 
The computational box has been rescaled for an improved viewing, and to attain the original aspect 
ratio, the box should be stretched 10 times along the axial direction. The magnetic topology for the 
total field is quite simple, as the lines appear  slightly bent. 

Figure~\ref{fig4}  (\emph{bottom row}) also shows a view from the side and the top of 
the 3D current sheets at time $550\, \tau_A$. The current sheets, elongated along the 
axial direction, look space filling when watched from the side of the computational box, 
but the view from the top shows that the filling factor is actually small, as they are 
almost 2D structures.

So far we have analyzed the topology of the field-lines only in the mid-plane $z=5$.
In Figure~\ref{fig5} we show the current density and magnetic field lines of $\mathbf{b}_{_\perp}$
in the mid-plane ($z=5$) and in two other $x$-$y$ planes close to the boundaries
($z=1$ and $9$, the axial length $L=10$). The behavior is similar at different heights although in the 
plane $z=9$, closer to the forced boundary $z=10$, the topology of the field appears to be affected 
to some extent by the sheared velocity forcing (\ref{eq:f0}) directed along the $y$~direction.
The field-lines in fact show a small alignment directed along $y$ close to the boundary $z=10$.

\begin{figure}
\plotone{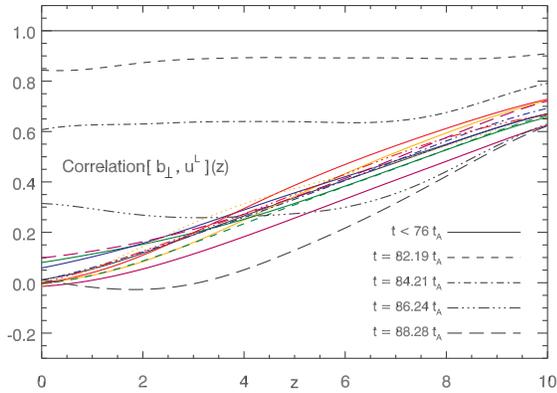}
\caption{Correlation [eq.(\ref{eq:cor})] between the magnetic field $\mathbf{b_{_\perp}}$
 and the boundary forcing velocity $\mathbf{u^L}$ 
 [applied at the boundary $z=10$, see (\ref{eq:f0})] as a function
 of the axial coordinate $z$ at selected times.
 Correlation is equal to 1 during the linear stage ($t<76\, \tau_A$) and decreases
 as the system transitions to the nonlinear stage ($82.19\, \tau_A \le t \le  88.28\, \tau_A$).
 The different colors show the correlation during the fully nonlinear stage at
 10 different times separated by $\Delta t = 40\, \tau_A$
 in the interval $200\, \tau_A \le t \le 600\, \tau_A$.
\label{fig6}}
\end{figure}

The influence of the boundary forcing over the magnetic field can be expressed 
\emph{quantitatively} through the correlation between the magnetic field $\mathbf{b_{_\perp}}$
in the plane $z$ and the boundary forcing velocity $\mathbf{u^L}$ [eq.(\ref{eq:f0})]:
\begin{equation} \label{eq:cor}
\mathrm{Cor} \left[ \mathbf{b}_{_{\!\perp}}, \mathbf{u^L} \right] \left( z \right) =
\frac{   \int \! \!\!  \int \!  \mathrm{d}x \mathrm{d}y \, \mathbf{b_{_\perp}} \! \cdot \mathbf{u^L} }
{  \Big[ \int \! \!\!  \int \!  \mathrm{d}x \mathrm{d}y \, \mathbf{b^2_{_\perp}} \cdot 
            \int \! \!\!  \int \!  \mathrm{d}x \mathrm{d}y \,  (\mathbf{u^L})^2 \Big]^{\frac{1}{2}} }
\end{equation}
In Figure~\ref{fig6} we plot the correlation as a function of the axial coordinate $z$
at selected times. In the linear stage, until time slightly bigger than $t = 76\, \tau_A$
whereafter the system transitions to the nonlinear stage (Figures~\ref{fig1}, \ref{fig2}, 
and \ref{fig3}), the magnetic field is a mapping of the boundary velocity
therefore as expected the correlation is equal to 1.
In fact for the simulation presented in this section (run~A), for which we have imposed 
the shear velocity profile (\ref{eq:f0}) at the top plate $z=10$ and a vanishing velocity
at the bottom plate $z=0$, the magnetic field in the linear stage is given by
eq.~(\ref{eq:lin1s}) [or (\ref{eq:diff1}) with $\mathbf{u^0}=0$ including diffusion] 
therefore the correlation is 1 as $\mathbf{b_{\perp}}$ is proportional to the boundary 
velocity $\mathbf{u^L}$.

Next as the system transitions to the nonlinear stage releasing most of the accumulated 
magnetic energy the correlation between the magnetic field and the boundary
forcing velocity decreases swiftly, at a faster pace the farther from the forced boundary
$z=10$, as shown by the curves at times $82.19 \le t / \tau_A \le 88.28$.

The correlation during the fully nonlinear stage is shown with color lines 
at $10$ selected times separated by $\Delta\, t = 40\, \tau_A$
in the interval $200\, \tau_A \le t \le 600\, \tau_A$. 
The correlation vanishes near the bottom boundary and then  grows almost 
linearly with $z$ up to $\sim 0.6$ at the top boundary. As expected the 
correlation is bigger near the forced boundary and fades towards the interior 
of the computational box. The magnetic field is overall weakly correlated 
with the forcing velocity, and at most reaches a mild correlation close to 
the forcing boundary, but it is never close to a strong correlation $Cor = 1$.

\begin{figure}
\plotone{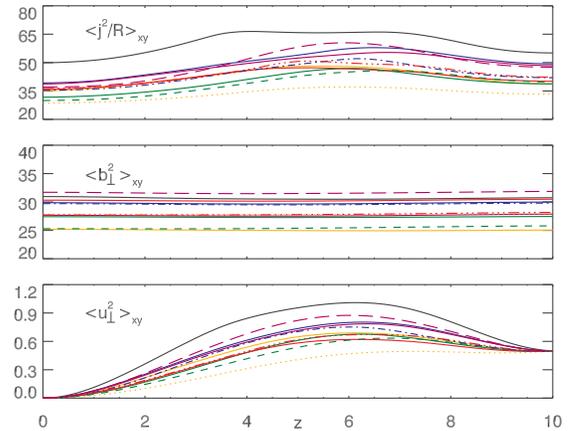}
\caption{Two-dimensional averages in the x-y planes of the ohmic dissipation 
$j^2/Re$, the magnetic field $\mathbf{b_{_\perp}^2}$, and the velocity field
$\mathbf{u_{_\perp}^2}$, as a function of the axial coordinate $z$. 
The different colors represent 10 different times separated by $\Delta\, t = 40\, \tau_A$
 in the interval $200\, \tau_A \le t \le 600\, \tau_A$.
\label{fig7}}
\end{figure}

In Figure~\ref{fig7}  we show the 2D averages in the x-y planes of the magnetic and 
velocity fields and of the ohmic dissipation $j^2/R$ plotted as a function of $z$ at 
different times. The behaviour is very similar to our previous simulations with different
(\emph{vortical}) forcing patterns \citep{rved08}. 
These macroscopic quantities are smooth and present almost no variation 
along the axial direction. The velocity must approach its boundary values at $z=0$ and $10$,
and in the volume grows to values higher than the boundary velocity $\mathbf{u_L}$.
In fact also the velocity inside the volume is not a mapping of the boundary forcing but 
develops self-consistently, in particular the plasma jets at reconnection locations
contribute too to its average.
The reason of the overall smooth behavior of these quantities
is that every disturbance or gradient along the axial 
direction is smoothed out by the fast 
propagation of Alfv\'en waves along this direction; their propagation time $\tau_A$ 
is in fact the fastest timescale present (in particular faster than the nonlinear timescale
$\tau_A < \tau_{nl}$), and then the 
system tends to be homogeneous along this direction.

\subsection{Dissipation versus Reynolds Number and Transition to Turbulence}

\begin{figure}
\plotone{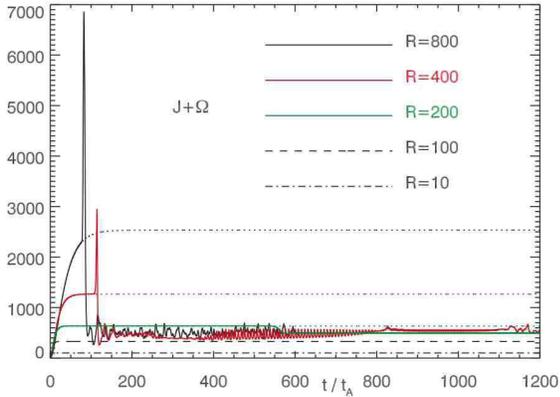}
\caption{Total (Ohmic plus viscous) dissipation rates as a 
function of time for different Reynolds numbers. 
The dashed curves are the linear saturation curves for
the ohmic dissipation rates  [eq.~(\ref{eq:diff3})] for the different cases. 
The overlap for $Re \ge 200$ suggests that total dissipation is independent
from Reynolds number at higher values.
\label{fig8}}
\end{figure}

Simulations~A-E differ only for the value of the Reynolds number
(and corresponding grid resolution), all other parameters are the same
including the amplitude of the perturbations. In Figure~\ref{fig8} we plot the total
dissipation for the 5 simulations as a function of time.

The simulation described in the previous paragraph had $Re = 800$,
after the first big dissipative peak its signal displays a complex temporal
structure that arises from the underlying turbulent dynamics.
Decreasing the value of the Reynolds number to $Re = 400$ the structure
of the signal has a simpler structure, with an almost sinusoidal
form of different amplitude and period $\sim 8 \tau_A$ in some time intervals.

The dotted lines represent the linear diffusive behavior described in \S~\ref{par3}.
These solutions [eq.~(\ref{eq:diff3})] are obtained when the nonlinear terms can
be neglected. This can happen either because they vanish exactly
due to the symmetry of the forcing velocity (as discussed in \S~\ref{par3})
or equivalently \emph{when nonlinear terms are depleted by diffusion},
i.e.\ when \emph{diffusion dominates the dynamics}.

At $Re = 200$ diffusion affects the system substantially,  the instability
takes longer to develop, and  afterward the actual signal and linear
saturation curve differ little. There is no first big dissipative peak, for a long
time up to $t \sim 550\, \tau_A$ the signal follows the linear saturation 
curve~ (\ref{eq:diff3}). Afterwards it departs slightly displaying a sinusoidal
behavior of very small amplitude.

At lower Reynolds number ($Re=100$ and $10$ are shown) 
diffusion dominates the dynamics, and no small scales
is formed, while ohmic dissipation and energy follow the 
curve~(\ref{eq:diff2})-(\ref{eq:diff3}) describing the diffusive equilibrium that is formed,
as nonlinear terms are completely depleted.

It is very interesting to notice that total dissipation for simulations
A, B and C with $Re=800$, $400$ and $200$ substantially 
overlap each other.  
As typical with spectral numerical codes, which use Fourier transforms to compute
derivatives, dissipation due to numerical \emph{implicit} diffusion is very small. 
We have checked that the energy conservation equation,
which can be derived from  eqs.~(\ref{eq:eq1})-(\ref{eq:eq3}),
is numerically verified within a very small error:  it is around $2\%$ 
for the lower resolution simulations and decreases below $0.5\%$ at 
higher resolutions \citep[see][Figure~5.5]{rapp06}. 
Therefore in the simulations presented
here the dissipation rates are substantially due to the
\emph{explicit dissipative terms} present in 
equations~(\ref{eq:eq1})-(\ref{eq:eq2}), while the diffusion
due to the numerical schemes is negligible.

We had already seen this behavior in
our previous simulations with vortex forcing \citep{rved08},
but in that case the boundary velocity was slightly different in each 
simulation. In fact it was built by a linear combination of Fourier
modes with \emph{random} amplitudes normalized to have the same rms,
but the spatial pattern was different for every simulation as determined
by the random amplitudes. In the simulations presented here 
the forcing is exactly the same for each simulation, as it is simply
a single Fourier mode [eq.~(\ref{eq:f0})].

As the forcing is the same for each simulation the overlap is more evident, 
and makes stronger the claim that
total dissipation is \emph{independent of the Reynolds number beyond
a threshold}. We conjectured this hypothesis \citep{rved08} because
the energy flux entering the system (Poynting flux) and the energy
transport from large to small scales due to the development of a turbulent
dynamics are both independent from the Reynolds number, where the turbulent
transport exhibits this property only for a \emph{sufficiently high value} of the Reynolds number.

This unfortunately \emph{does not} imply at all that the thermodynamical and
radiative outcome is independent of the Reynolds number. In fact how field-lines 
are heated strongly depends on the dynamics and properties around the
single current sheet. These become thinner and thinner at higher Reynolds numbers
and the overall dynamics more chaotic. Hence the thermodynamical properties
of the field-lines that cross the current sheets (elongated along the 
axial direction) and get so heated \emph{impulsively} are all to be explored. 

An open question is whether the dissipation is independent of the Reynolds
number also in the single current sheets, or their number and properties
change to attain independence for the total dissipation. Research in
this area is active \citep{lsc07,lap08,ser09, slu09,csd09,bhyr09,lus09,hb10,ser10} 
and has already shown that at relatively high
Reynolds numbers reconnection departs the classic Sweet-Parker scalings
in the MHD regime.
Furthermore for a plasma in coronal conditions kinetic effects cannot be excluded
a priori, in fact they might play an important role in dissipating energy through
particle acceleration. Also the high value of the magnetic Prandtl number could have
a bearing \citep{she04}. Nonetheless the total dissipation, integrated over the whole volume,
is likely to not depend on the detailed small-scale dissipative mechanism. In fact the energy
injection rate~[eq.~(\ref{eq:pf})] depends only on the large scale fields and
the transfer of this energy  toward the small scales appears to be local \citep{rv10},
i.e.\ it is determined by the fields at neighboring large scales, thus making also
the transfer energy rate toward the small scales independent of the Reynolds
number (provided it is beyond a minimal threshold to have scale separation).

\begin{figure}
\plotone{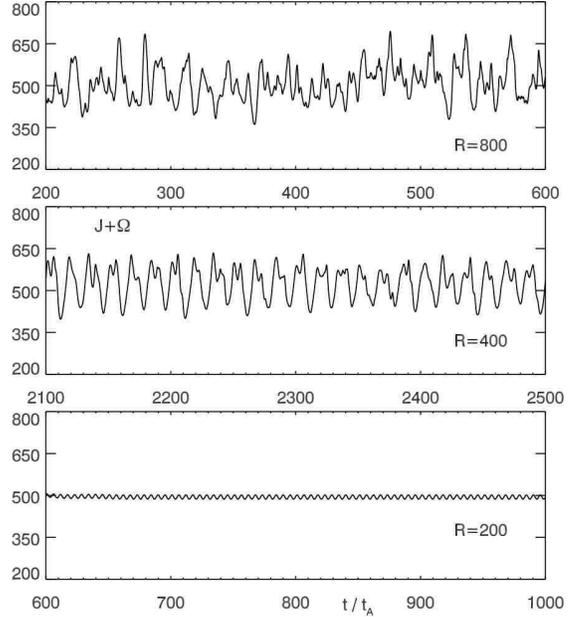}
\caption{Transition to turbulence, total ohmic and viscous dissipation
as a function of time form simulations A, B and C (displayed on the same scale for
an equal interval of time).
All the simulations implement $c_A = 200$, but different Reynolds numbers,
from $Re = 200$ up to $800$. For Reynolds numbers lower than $100$, the
signal is completely flat following exactly the linear saturation curve eq.~(\ref{eq:diff3}). 
At higher Reynolds numbers, smaller temporal structures are present displaying 
a transition to turbulence.
\label{fig9}}
\end{figure}

In Figure~\ref{fig9} we show a close-up of total dissipation for an equal interval
of time $\Delta\, t = 600\, \tau_A$ on the same scale in order to highlight their temporal
structures.
At higher Reynolds number smaller time frequencies
are present, a clear indication of a transition to turbulence \citep{fr95}.
For Reynolds numbers lower than $100$, the
signal is completely flat following exactly the linear saturation curve eq.~(\ref{eq:diff3}).

\subsection{Spectral properties}

In order to study the spectral properties of the system and compare them with
those of previous simulations we have performed a new simulation
where the sheared forcing is applied on \emph{both the top and bottom
plates} (run~F) with reversed direction [eqs.~(\ref{eq:f0}) and (\ref{eq:f2})].
We compare these results with those obtained from a previous simulation
(run~G), that has all the same parameters, except that on the two boundary planes
a large-scale ``vortex-like'' velocity pattern with the same rms 
($\langle \mathbf{u^2} \rangle = 1/2$) is applied.
Both simulations have been performed with a numerical resolution of
$512\times 512\times 200$ grid points, and hyperdiffusion with dissipativaty $n=4$ and
$R_4 = 10^{19}$.

\begin{figure}
\plotone{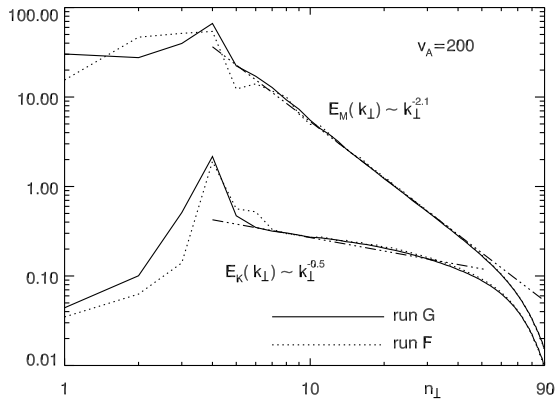}
\caption{Kinetic ($E_K$) and Magnetic ($E_M$) Energy spectra as a function
of the orthogonal  wavenumber $n_{_\perp}$ for simulations~F and G.
The spectra are very similar whether a vortex-like or shear velocity pattern
stirs the footpoints of the magnetic field-lines.
\label{fig10}}
\end{figure}

In Figure~\ref{fig10} we plot the energy spectra obtained from runs~F and G.
They are very similar to each other. Both the velocity and magnetic fields
develop inertial ranges following power laws, and overlap
each other.
For both runs the spectral index of the kinetic spectrum ($\sim -0.5$) is much smaller
than that of the magnetic energy ($\sim -2.1$), that is steeper than 
kolmogorov ($-5/3$). Also the kinetic energy has a lower value than the
magnetic energy, as already noticed for the integrated quantities (Figure~\ref{fig1}).

The energy that is injected into the system for unit time is the integrated
Poynting flux
\begin{equation} \label{eq:pf}
S = c_A \int_{z=L} \mathrm{d}a (\mathbf{u_{_\perp}^L} \cdot \mathbf{b_{_\perp}})
   -   c_A \int_{z=0} \mathrm{d}a (\mathbf{u_{_\perp}^0} \cdot \mathbf{b_{_\perp}})
\end{equation}
where $\mathbf{u_{_\perp}^L}$ and $\mathbf{u_{_\perp}^0}$ are the imposed
velocity patterns at the top and bottom planes.

For run~G, we excite all wavenumbers $3 \le n_{_\perp} \le 4$, while for run~F we
\emph{excite only one Fourier component} as
$\mathbf{u_{_\perp}^L} = - \mathbf{u_{_\perp}^0} = 
\sin \left( 8\pi x + 1 \right) \mathbf{\hat{e}_y}$, i.e.\ we are injecting energy in the
system only at $\mathbf{n_{in}} = 4 \cdot 2\pi\, \mathbf{\hat{e}_x}$, the wavenumber
$4$ along $x$. This can be noticed also in Figure~\ref{fig10}, where the kinetic spectrum
for run~G at $n=3$ is higher than for run~F, as part of the energy is injected also at
$n=3$ in the vortical case.

The lower level for the kinetic spectrum is due to the boundary conditions that roughly
set the value or the velocity at the injection wavenumbers inside the volume.
In the simple linear case this is given by eq.~(\ref{eq:lin2}). In the shear case
we would have $E_K(4)= 1/2 \cdot V \cdot \langle \mathbf{u^2} \rangle = 2.5$ ($V=10$ is the volume)
in the linear regime, and from Figure~\ref{fig10} we
notice that also in the nonlinear regime $E_K(4) \sim 2.5$. On the other hand the
magnetic field grows linearly in time [eq.~(\ref{eq:lin1})] until a balance  is reached
between the energy flux that is injected at this scale and the flux of energy
flowing towards smaller scales through a turbulent cascade.

The magnetic energy spectra of the two simulations are slightly different at the large
scales with $n_{_\perp} \le 5$. The large scale dynamics is in fact slightly different in the two cases.

In the vortical case (run~G) energy is injected in all modes with wavenumbers 
$3 \le n_{_\perp} \le 4$ that then cascades toward smaller scales.
In the shear case (run~F) energy is injected only at one wavenumber $\mathbf{n_{_\perp}} = (4, 0)$.
We have already noticed in \S~\ref{sec411} that although we continue shearing the footpoints
of the field-lines with our 1D forcing [eq.~(\ref{eq:f0})] in the nonlinear stage the
orthogonal magnetic field is organized in magnetic islands (Figure~\ref{fig3}), so that
it is no longer a mapping of the boundary velocity. Although energy is injected only in the
wavenumber 4 along x, energy is then redistributed by the nonlinear terms also
to modes with wavenumbers along y at the large scales, and a small inverse cascade
is present as in run~G.
This is the basic mechanism by which magnetic islands are sustained throughout
the simulation in the nonlinear stage.

\section{Conclusions and Discussion} \label{sec:con}

In this paper we have investigated the dynamics of the Parker problem for the 
heating of coronal loops when the footpoints of the magnetic field-lines are
stirred by a 1D shear velocity pattern at the photosphere-mimicking boundary, 
and compared these results with those previously obtained when a more complex 
``vortex-like'' velocity pattern was imposed \citep{rved08}. This very simple
forcing is ideal to investigate the origin of turbulence in coronal loops
and the influence of the boundary velocity forcing on the dynamics of the system.

We will also compare our results with those of \cite{hp92} and of the more recent 
simulations of \cite{dka05,dlkn09}.

In summary, the main results presented in this paper are the following:
\newcounter{ctr}
\begin{list}{\arabic{ctr}.}{\leftmargin=1.2em}
\usecounter{ctr}
\item Initially the sheared velocity forcing induces a sheared perpendicular magnetic field
inside the volume. The resulting current layers are known to be unstable to tearing modes
\citep{fkr63}. In fact when the system transitions from the linear to the nonlinear stage
it is due to a multiple tearing instability, as shown in Figure~\ref{fig3}.
But once the system has become fully nonlinear the dynamics are fundamentally different.
As the nonlinear terms no longer vanish they now do transport energy from the large
to the small scales where in correspondence of the X-points \emph{nonlinear} magnetic reconnection
takes place, without going through a series of equilibria disrupted by tearing-like
instabilities. 
Similarly to the case with disordered vortical boundary forcing velocities
\citep{rved07, rved08} in the fully nonlinear stage the system is highly dynamical and
chaotic (and increasingly so at higher Reynolds numbers). For this we do not
observe secondary tearing of the current sheets as in 2D high-resolution simulations 
of decaying MHD turbulence \citep{bw89}, as now at the small scales  
fast turbulent dynamics take place.

\item The dynamics of the Parker model do not depend strongly on the pattern of the velocity
forcing that mimics photospheric motions, 
as far as they are constant in time (we defer the investigation
of time-dependent boundary forcing to a future work).
The shear forcing [eq.~(\ref{eq:f0})] applied only at the top plate
is a very simple and ordered one-dimensional  forcing. We have shown that the resulting dynamics are 
very similar to those developed when a more complex and disordered
``vortex-type'' forcing  velocity is applied. We conclude that the turbulent properties 
 of the system are not induced by the \emph{complexity} of the path that the footpoints
follow. It is rather the system itself to be intrinsically turbulent, and turbulence develops as we continuously
inject energy at the scale of photospheric motions ($\sim 1,000\, km$).

\item The system reaches a statistically
steady state where, although the footpoints of the field-lines are continuously dragged by
the forcing  shear, this does not induce a sheared magnetic field in the computational box.
In fact the topology of \emph{the magnetic field is not a mapping of the forcing velocity field}.
Nonlinear interactions are able to redistribute the energy that is injected only at the 
wavenumber $\mathbf{n_{_\perp}} = 4\, \mathbf{\hat{e}_x}$ also to perpendicular wavenumbers
and to smaller wavenumbers through an MHD turbulent cascade.
In physical space this corresponds to the magnetic field being organized in magnetic islands,
to small-scales formation (current sheets elongated along the axial direction)
and to magnetic reconnection taking place.

\item Kinetic and Magnetic energies develop an inertial range where spectra exhibit a power-law behavior.
Fluctuating magnetic energy dominates over kinetic energy. Spectra and integrated quantities, like energies
and total dissipative rates, have values similar to those obtained with a vortex-type forcing velocity.
In particular the total dissipation rate of the same system simulated with different Reynolds number
appear to overlap each other beyond $Re = 200$, suggesting that this is independent of the Reynolds
number beyond a threshold.
\end{list}

As shown in Figures~\ref{fig1}, \ref{fig2} and \ref{fig3} initially the system until time
$t \sim 79\, \tau_A$ follows the linear curves~(\ref{eq:lin2}) and (\ref{eq:diff1}).
Up to this point the shear velocity at the top boundary induces a sheared magnetic field in the volume.
As discussed in \S~\ref{sec:runa} we have introduced a perturbation mimicking those naturally
present in the corona. With no perturbation the system would
relax over the resistive diffusive timescale $\tau_R$ ($\sim 25\, \tau_A$ for run~A)
in a saturated diffusive equilibrium as described in (\ref{eq:lin2}) and (\ref{eq:diff1}). 
While the simulation presented here used a very small amplitude for 
the perturbation ($\epsilon = 10^{-16}$), we have performed shortest simulations with different values
for the amplitude. As expected for higher values of $\epsilon$ the instability develops sooner and 
for smaller values later, always following the linear curves until the instability transitions
to the nonlinear stage.
The more complete and systematic analysis of \cite{rom04,rom09} in 2D  confirms this behavior.

\cite{dlkn09} have performed a similar simulation with a lower resolution and with a fixed value 
for the perturbation and for a time interval
that covers only the initial stage of our simulations. They in fact stop right after the first big dissipative peak,
that in our Figures~\ref{fig1}, \ref{fig2} and \ref{fig3} corresponds at $t \sim 100\, \tau_A$.

Continuing the simulation, and using a higher numerical resolution, the system reaches a 
statistically steady state where magnetic energy
consistently fluctuates around a mean value and the shear is not recreated in the topology of the
orthogonal magnetic field. Their analysis is then limited to a \emph{transient} event taking place only
during the early stages of the dynamics, and that afterward does not repeat. 

As shown in \cite{rved08} during the linear stage the system is able to accumulate energy well beyond
the average value maintained in the nonlinear stage only if the boundary forcing velocity satisfies
the condition that its \emph{vorticity is constant along the streamlines}. The sheared profiles used
in this paper satisfy this condition as well the profile used by \cite{dlkn09} (a linear combination
of 6 sheared profiles). 

These profiles are a very small subset of all the possible forcing profiles, and while they are very
useful to get insight into the origin of turbulence in coronal loops they are not representative
of the disordered photospheric motions, for which the strong stress buildup required for 
secondary instability to develop does not take place.  
The significance of their conclusions is then strongly diminished.

Furthermore the Parker angle for this system cannot be defined as the relative angle between magnetic
field-lines at which the system becomes unstable.
This is not a \emph{well-posed} definition.
In fact for given initial conditions the angle or equivalently the time (as the linear equation 
(\ref{eq:lin2}) and (\ref{eq:diff1}) imply) at which the instability develops depend on the value 
of the amplitude of the perturbation that we add to the system. 
Depending on the value of the perturbation the Parker angle so defined \emph{is not unique}.

On the other hand in the fully nonlinear stage the average magnetic field line magnitude
fluctuates around a mean value.
It is then possible to give a unique value for the Parker angle, defined now as the average 
inclination of the magnetic field-lines respect to the axial direction as done in \cite{rved07,rved08},
and as originally introduced by \cite{park88}.

As summarized in  \S~\ref{sec:ed}  the one-point closure model developed by \cite{hp92}
splits the domain into large and small scales.
They conjecture that the large-scale fields evolve into a stationary laminar regime, the field magnitudes
determined by the effective diffusion coefficients. These laminar regimes correspond to our linear saturated
diffusive regimes computed in \S~\ref{par3}. In Figure~\ref{fig8} the dotted lines show such
diffusive curves for different values of the Reynolds numbers.  

In their model the large-scale fields computed in this way are used
to obtain $S$, the energy flowing into the system for unit time at the boundary (the power)
due do the work done by photospheric motions on the magnetic field-lines footpoints. 
They also calculate, through an EDQNM approximation,
the value of the spectral energy flux $\epsilon$ flowing along the inertial range at the small scales. 
Both $S$ and $\epsilon$ are functions of the effective diffusion coefficients, and the solution 
of the problem results requiring balance between the two powers  $S = \epsilon$ 
[$S$ and $\epsilon$ have both the dimension of a power, i.e.\ energy over time, as
$S$ is the Poynting flux integrated over the boundary surface and $\epsilon$ is integrated
over the whole volume as in \cite{rved07}].

As shown in our simulations the large-scale fields are not laminar, and they are stationary
only statistically. Nevertheless it is useful to use \cite{hp92} model in order to understand
why it is not applicable. From Figure~\ref{fig8} we can estimate that the \emph{effective} Reynolds
number for which the diffusive regime dissipation matches the dissipation of the simulated system is
$R_{eff} = 150$. Unfortunately this values is too low for their model to work.
In fact for $R = 150$ the dynamics are so diffusive that only a few modes of the order of the injection
scale ($\sim 1,000\, km$) are not suppressed but only reduced in magnitude. Therefore
there is no flux of energy at the small scales $\epsilon = 0$.

At a more fundamental level the idea to split the domain into large and small scales
does not work because nonlinearity cannot be confined only at the small-scales.
As shown by our simulations \emph{nonlinearity} is at work at all scales, and unfortunately 
this fundamental aspect \emph{cannot be circumvented}.

Finally the use of RMHD equations is valid as far as the magnetic 
field fluctuations $\mathbf{b_{_\perp}}$ are small compared to the
axial magnetic field $B_0$. This seems particularly apt to describe
the dynamics of long-lived slender loops that apparently show no
dynamics while shining bright at the resolution  scale ($\sim 800\, km$) 
of current state-of-the-art  X-ray and EUV imagers onboard 
Hinode and Stereo. Clearly these results do not apply to highly
dynamical active regions where dynamics cannot be modeled
as fluctuations about an equilibrium configuration.

The series of simulations that we have performed proves 
that dragging the footpoints of magnetic field-lines in the 
Parker problem quickly triggers nonlinear dynamics for small
values of the orthogonal magnetic fields, and that these small
magnetic field fluctuations are able to transport a considerable
amount of energy toward the small scales with the overall
energy flux $\sim 1.6 \times 10^6\, erg\, cm^{-2}\, s^{-1}$ 
\citep{rved08} in the lower range of the observed constraint
$\sim 10^7\, erg\, cm^{-2}\, s^{-1}$.

This prevents the orthogonal magnetic fluctuations
to grow to an arbitrarily high value, self-consistently limiting
the dynamics of the Parker problem to small fluctuations
if the initial conditions are given by a  uniform strong
axial magnetic field.

\acknowledgments

We thank Russ Dahlburg (NRL) for useful discussions. 
A.F.R.\ was supported by the NASA Postdoctoral Program. 
This research was supported in part by the Jet Propulsion Laboratory, 
California Institute of Technology under contract with NASA, 
and in part by ASI contract n.~I/015/07/0 Exploration of the Solar System.
Financial support by the European Commission through the SOLAIRE Network
(MTRN-CT-2006-035484) and by the Spanish Ministry of Research and
Innovation through projects AYA2007-66502 and CSD2007-00050 is 
gratefully acknowledged.
Simulations have been performed through the NASA Advanced 
Supercomputing SMD award 09-1112 and at CINECA (Italy). 
A.F.R.\ thanks the Leverhulme Trust International Network for 
Magnetized Plasma Turbulence for travel support.

\end{document}